\documentclass[12pt,a4paper]{article}

\usepackage[T1]{fontenc}
\usepackage[utf8]{inputenc}
\usepackage{float}
\usepackage{graphicx}
\usepackage{array}     
\usepackage{multirow}  
\usepackage{subcaption} 
\usepackage[thinlines]{easytable}
\usepackage{hyperref}
\usepackage{wrapfig}
\usepackage{dingbat}
\usepackage{color}
\usepackage{refcount}
\usepackage[table]{xcolor}
\usepackage[toc,page]{appendix}
\usepackage{bbm}
\usepackage{enumitem}
\usepackage{amsmath,amssymb,amsthm}
\usepackage{mathtools}
\usepackage[english]{babel}
\usepackage{natbib}
\usepackage{xcolor}
\usepackage[normalem]{ulem}
\usepackage{booktabs}  
\usepackage{multirow}  
\usepackage{graphicx}  
\usepackage{bibunits}
\mathtoolsset{showonlyrefs}

\renewcommand{\baselinestretch}{1.15}

\theoremstyle{definition}
\newtheorem{theo}{Theorem}[section]
\newtheorem{lemma}[theo]{Lemma}
\newtheorem{prop}[theo]{Proposition}
\newtheorem{cor}[theo]{Corollary}

\newtheorem{algorithm}[theo]{Algorithm}
\newtheorem{rem}[theo]{Remark}
\newtheorem{ass}[theo]{Assumption}

\numberwithin{equation}{section}

\definecolor{violet}{RGB}{148,0,211}
\definecolor{olivegreen}{RGB}{107,142,35}

\newcommand{\R}{\mathbb{R}}

\newcommand{\E}{\mathbb{E}}
\newcommand{\N}{\mathbb{N}}

\usepackage{algorithm}
\usepackage{algpseudocode}

\usepackage{authblk}
\usepackage{titling}

\date{} 
\begin{document}

\title{\textbf{Monitoring Time Series for Relevant Changes}}

\author{
Patrick Bastian$^{1}$, Tim Kutta$^{2,*}$, Rupsa Basu$^{3}$, Holger Dette$^{1}$ \\
\\
$^{1}$Fakultät für Mathematik, Ruhr-Universität Bochum, Germany \\
\texttt{\{patrick.bastian, holger.dette\}@rub.de} \\
$^{2}$Department of Mathematics, Aarhus University, Denmark \\
\texttt{tim.kutta@math.au.dk} \\
$^{3}$Institute for Econometrics and Statistics, Universität zu Köln, Germany \\
\texttt{rbasu@uni-koeln.de}
}




\maketitle
$ $\\[-10ex]
\begin{abstract}
We consider the problem of sequentially testing for changes in the mean parameter of a time series, compared to a benchmark period. Most tests in the literature focus on the null hypothesis of a constant mean versus the alternative of a single change at an unknown time. Yet in many applications it is unrealistic that no change occurs at all, or that after one change the time series remains stationary forever. We introduce a new setup, modeling the sequence of means as a piecewise constant function with arbitrarily many changes. Instead of testing for a change, we ask whether the evolving sequence of means, say $(\mu_n)_{n \geq 1}$, stays within a narrow corridor around its initial value, that is, $\mu_n \in [\mu_1-\Delta, \mu_1+\Delta]$ for all $n \ge 1$. Combining elements from multiple change point detection with a Hölder-type monitoring procedure, we develop a new online monitoring tool. A key challenge in both construction and proof of validity is that the risk of committing a type-I error after any time $n$ fundamentally depends on the unknown future of the time series. Simulations support our theoretical results and we present two real-world applications: (1) healthcare monitoring, with a focus on blood glucose tracking, and (2) political consensus analysis via citizen opinion polls.\let\thefootnote\relax   
\footnote{* Corresponding author.}
\let\thefootnote\arabic{footnote}  

\end{abstract}


\defaultbibliographystyle{apalike}
\defaultbibliography{reference}

\begin{bibunit}
\section{Introduction}

 In this paper, we consider the problem of change point detection in the sequence of means of a  real-valued time series $(X_n)_{n \in \mathbb{N}}$. At the outset of the procedure, a \textit{stable training sample} of size $N$ is available to the analyst, where $\mathbb{E} X_1=\mathbb{E} X_2=\cdots =\mathbb{E} X_N$. 
The training sample serves as a benchmark for the subsequent inference and theoretical results are formulated for $N \to \infty$. After time $N$, the mean evolves dynamically and is described by 
\begin{align}\label{e:defpsi} 
\mu_n :=  \mathbb{E} [X_{n}] = \Psi\Big(\frac{n}{N}\Big), \qquad n \in \mathbb{N},   \end{align}
where 
$\Psi: (0, \infty) \to \mathbb{R}$ is 
 a  piecewise constant function, that is 
\begin{align}\label{e:defpsinew} 
\Psi(x) := \sum_{i\ge 1} \psi^{(i)} \cdot \mathbb{I}\{\theta^{(i-1)}< x \le \theta^{(i)}\},
\end{align}
$\theta^{(0)}=0$, $1  < \theta^{(1)}  < \theta^{(2)} < \ldots $ and $\psi^{(1)}= \mu_1$. After time $N$, the training period is over and the \textit{monitoring phase} begins. Here, data arrive sequentially and statistical inference is based on the increasing samples 
\[
\{X_1,\cdots ,X_{N+1}\},\{X_1,\cdots ,X_{N+2}\}\cdots ,\{X_1,\cdots ,X_{N+n}\},\cdots  
\]
for an indefinite (and theoretically infinite) time span. We are interested in monitoring the time series for \textit{approximate mean stability}, where the mean $\mu_n$ is always close to the mean of the training sample. We can formalize this as a decision  problem as follows: for a threshold $\Delta>0$, we are interested in testing the hypotheses pair
\begin{align}
    H_0(\Delta): \sup_{1 \le x <\infty}|\Psi(1)-\Psi(x)|\le \Delta, \quad \textnormal{vs.}\quad H_1(\Delta): \sup_{1 \le x <\infty}|\Psi(1)-\Psi(x)|>\Delta\label{e:H0}
\end{align}
and refer to this as {\it $\Delta$-approximate mean stability}, if the null hypothesis is satisfied.
\medskip

\textbf{\hypertarget{myword}{Comparison to sequential change point detection}} The problem of sequentially searching for deviations in a model parameter compared to the training period has been studied in a number of works. However, typically, in such models the function $\Psi$ takes the form of a Heaviside function 
\[
\Psi(x):= \mu \cdot\mathbb{I}\{x \le \theta\}+\nu \cdot\mathbb{I}\{x > \theta\},
\]
with $\theta>1$ and $\nu \in \mathbb{R}$. In particular there exists at most one change.
The task is then to sequentially test whether $\nu=0$ (no change) or $\nu\neq 0$ (one change), which corresponds to the hypotheses pair \eqref{e:H0} with $\Delta=0$ and the above choice of $\Psi$.
Notice that $\Delta=0$, at least formally, requires the time series $(X_n)_{n \in \mathbb{N}}$ to be \textit{exactly mean-stationary forever}. However, in practice, there are always some non-stationarities 
 present. Accordingly, it is \textit{a priori} clear that $H_0(0)$ is at some point violated in the monitoring period and a test for $H_0(0)$ will reject eventually. 
Importantly, a rejection of $H_0 (0)$ indicates statistical significance but not necessarily scientific relevance of the detected violation of mean stationarity.
We provide a short example to illustrate this point.

\subsection{Motivating example: monitoring blood glucose levels} \label{sec_11}

\begin{figure}[ht]
    \includegraphics[width=1\linewidth, height= 8cm]{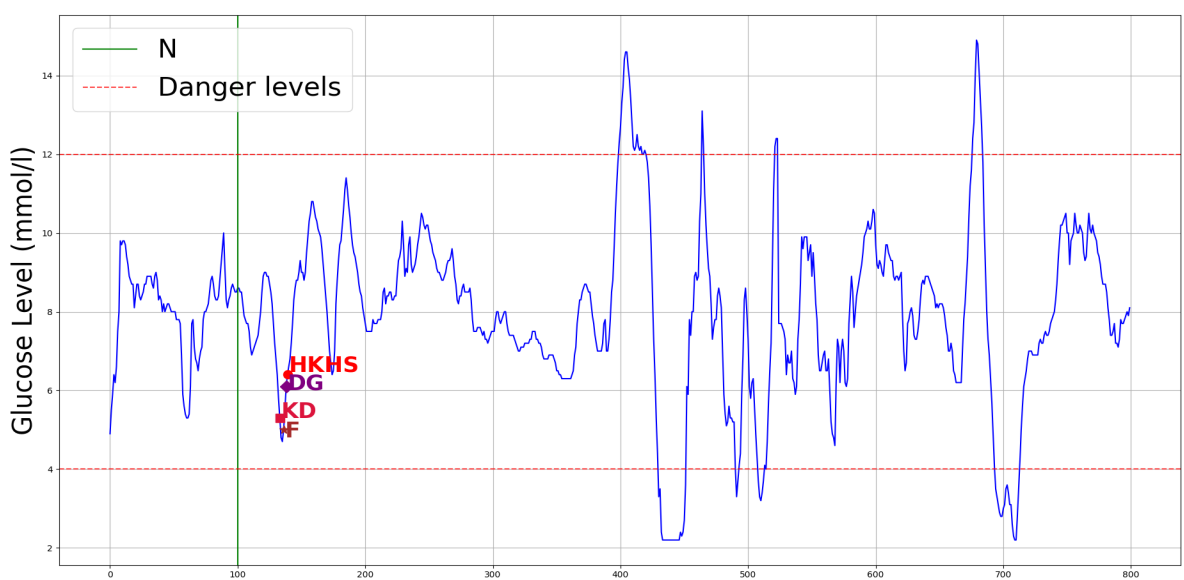}
    \caption{\it  
    Blood  glucose concentration
over time (blue) in millimoles per liter (mmol/L). Abbreviated annotations on the graph indicate change detection by standard monitoring methods. The  vertical line (green) marks the start of the monitoring period. Dashed horizontal lines (red) mark the boundaries of normal glucose levels.}
    \label{fig:glucose_levelMotivation} 
\end{figure}

This section highlights the advantages of monitoring approaches sensitive to meaningful variations (or relevant differences), demonstrated through the example of blood glucose tracking in diabetes mellitus. This example is based on an original dataset we collected from a volunteer with Type 1 Diabetes, who consented to the use of their data for this research and the publication of the associated analysis. For this individual, blood glucose concentrations were measured automatically with an average sampling frequency of five minutes over a total duration of 50 hours. Further details on the data collection process are provided in  Section \ref{App: diabetes example}.\

 To avoid negative health consequences associated with prolonged exposure to extreme glucose concentrations, diabetes patients require insulin doses tailored to their recent food intake. Blood glucose monitoring for insulin therapy treatment of type 1 diabetes supports not just  diagnostic purposes but also the daily life management of clients with impaired glucose metabolism. In addition, it facilitates identifying patterns in fluctuations based on factors like diet, exercise and medications (see \cite{mathew2020blood}).  Although individual targets may vary, we make our analysis quantitative by considering as a normative range values between 4 mmol/L and 12  mmol/L. These choices are motivated by the diagnostic criteria for  hypoglycemia (\cite{Cryer2009}) and diabetic ketoacidosis (\cite{Abbas2009}), respectively. Even when food intake and insulin supply are well managed, glucose levels are never constant (not even in healthy individuals). However, they should be approximately stable in the sense that they remain close to their long-term average and mostly stay within the above-mentioned boundaries.
Thus, we are in a setting where a real-valued time series is monitored, and only substantial deviations from its long-term mean  (large $\Delta$) are a cause for concern.\

The volunteer in this analysis uses a  closed-loop (Kaleido) insulin pump that automatically adjusts insulin delivery based on glucose levels for both basal (background) and bolus (mealtime). As a result, glucose concentrations for this particular dataset never remain at dangerous levels long enough to cause harm, and a successful monitoring procedure should therefore not trigger alarms unnecessarily. However, most existing procedures which look for changes for any $\Delta >0$ aim to detect any systematic shifts, no matter how small, and are thus often overly reactive.

To illustrate this oversensitivity of the commonly used methods, we apply some monitoring methods from \cite{horvath:kokoszka:huskova:steinebach:2003}, \cite{fremdt:2015}, \cite{dette:gossmann:2019}, and \cite{kutta:doernemann:2025}, hereafter referred to by their abbreviations HKHS, F, DG, and KD, respectively.  Figure \ref{fig:glucose_levelMotivation} displays the patient’s glucose concentration over time, with the green vertical line marking the start of the monitoring period  (we used  $N=100$ for the training sample, which
corresponds to $8.33$ hours). Note that the baseline period excludes any mealtimes and reflects glucose levels recorded overnight. We also indicate the first time each method detects a change, raises an alarm, and stops monitoring. All methods stop before even one observation exceeds the thresholds of the normative range. This clearly demonstrates that none of these methods are suitable for monitoring such a dynamic scenario, highlighting the need for different approaches for this type of data.

In Section \ref{App: diabetes example}, we analyze this glucose dataset using the new methods proposed in this paper and we briefly summarize the results: while our method is raising alarms for 
very small thresholds $\Delta > 0$, reasonable choices that are meaningful in practice (e.g., $\Delta=4$) result in no detections throughout the entire period. This is not due to low power - see the power analysis in Section \ref{sec_3} - but rather because our method is designed to distinguish relevant changes from insignificant fluctuations. In healthcare, as in many disciplines, an exaggerated response to minor fluctuations may result not only in increased financial burden but also in the unwarranted use of insulin therapy, potentially leading to avoidable side effects, see \cite{pouwer2009insulin}. It is the aim of this paper to provide a practical and reliable monitoring framework that focuses on raising alarms only for those changes that truly matter.


\subsection{Related literature} \label{sec_12}

There exists an enormous amount of  literature on change point analysis and in this brief review, we confine ourselves to the work which is most  closely related to our paper.
\smallskip

\textbf{Retrospective change point analysis} 
Change point analysis is often conducted retrospectively to identify structural breaks in a historical dataset. For comprehensive overviews of the change point literature, we refer the interested reader to the reviews of \cite{aue:horvath:2012,jandhyala:2013,Niuetal2016,truongetal2020} and \cite{cho:kirch:2024}. Most commonly studied in the literature are AMOC models, where a time series with \underline{a}t \underline{m}ost \underline{o}ne \underline{c}hange in a model parameter is permitted. In such models, the statistical tasks are testing for the presence of a change and then to estimate the change point location with high precision. 
AMOC models with tests for approximate parameter stationarity have been considered by \cite{dette:kokot:volgushev:2020} and \cite{Hu2025}. 

While AMOC models provide a simple way to model non-stationary time series, it is sometimes unrealistic that only one change should occur, especially over long stretches of time. As a consequence, a broad literature on multiple change point problems has developed,  where the task is to segment a non-stationary time series into stationary episodes. Informative reviews can be found in  \cite{cho:kirch:2024} and \cite{fryzlewicz:2024}. An important innovation in this field has been the transition from Binary Segmentation methods (see, e.g., \cite{vostrikova:1981,venkatraman:1992}) to Wild Binary Segmentation introduced by \cite{fryzlewicz:2014}. 
While most classical works on multiple change point detection are based on the assumptions of independent and normally distributed data, recently time series data has been explored. For instance, \cite{eichinger:kirch:2018} develop a MOSUM statistic for dependent time series. MOSUM methods are most appropriate where all changes are roughly of the same order of magnitude. If the magnitude of changes differ, they can be enhanced  by using multiple levels of resolution (see, e.g., \cite{messer:albert:schneider:2018}). A method that is specifically tailored to the detection of many, differently sized changes is the SMUCE method, proposed by \cite{frick:munk:sieling:2014} (see also \cite{dette:eckle:vetter:2020} for an extension of this approach to time series). The change point localization in these papers is related to our approach, which uses a multiscale change point detector as a subroutine.  However, our method rests on a Hölder-type statistic, which is quite different from  the one proposed by these authors. Moreover, in contrast to this literature, it aims for identifying practically significant change points. Optimality properties of multiple change point detection procedures have been studied by \cite{wang:yu:rinaldo:2020} and \cite{verzelen:fromont:lerasle:reynaudbouret:2023}   for independent subgaussian data. All this work aims at localizing  change points of arbitrarily small size. The only work we are aware of that aims for identifying practically relevant change points is \cite{bastian:basu:dette:2024}. These authors  consider  the retrospective multiple change point problem for  functional data and propose a new way to select from a set of multiple changes only those of large enough magnitude.  However, it  seems to be difficult to extend their method to online change point detection as considered in this paper.
\smallskip

\textbf{Sequential change point analysis} Retrospective methods provide diagnostic tools to investigate a complete, historical dataset. In contrast, sequential procedures aim for  the real-time analysis of an incoming data stream. There exist two main paradigms of sequential change point detection, namely that of \textit{Statistical Process Control} (SPC) and that of \textit{sequential testing}. These fields
have developed rather independently during the last decades.  One reason for this is
their different emphases when it comes to the sequential change point problem - besides presumably other
more mathematical reasons. The only comparison we are aware of is the recent work of \cite{Yu02102023},  which provides  a brief theoretical comparison of both approaches  in the univariate setting
with independent subgaussian observations and also indicates the mathematical challenges in such a comparison.

In SPC, one uses  models from classical sequential analysis (see \cite{siegmund:1985}) and  procedures are developed that detect a change shortly after it has occurred (short delay), while guaranteeing  a low frequency of false alarms if no change occurs (long average runtime). For an introduction to this paradigm, we refer the reader to the works of \cite{lai:1995,Woodall01101999} and \cite{lai:2001}. In this paper, we focus on the different paradigm of sequential testing, where procedures have to be sensitive to changes, while  controlling the probability of a false alarm over the entire monitoring period (i.e., control of the nominal level). For an overview of classical sequential testing we refer the reader to the recent review by \cite{aue:kirch:2024}. The original formulation of the testing problem dates back to the seminal work of \cite{chu:stinchcombe:white:1996}. Here, the authors consider monitoring a time series of linear models for a change in the slope parameter compared to a training period. In \cite{horvath:kokoszka:huskova:steinebach:2003}, a class of weighted change point detectors is presented for this problem, with a focus on fast detection of an early change. These statistics are referred to by \cite{aue:kirch:2024} simply as "CUSUM" statistics and they typically compare a parameter estimate from the training period with one of the (hitherto) observed monitoring period. 
Subsequent works, such as \cite{berkes:gombay:horvath:kokoszka:2004}, 
\cite{aue:horvath:huskova:kokoszka:2006}, \cite{aue:horvath:reimherr:2009} extend this approach to further models and even derive the asymptotic distribution of  the first time of rejection. Improvements of the CUSUM statistic 
have been proposed by \cite{fremdt:2015} and  \cite{dette:gossmann:2019} which address the fact that the CUSUM statistic  combines all available data from the monitoring period into a parameter estimate, even though some of this data comes from before the change. 
Modifications for high-dimensional and functional data respectively have been investigated by \cite{gosmann:stoehr:heiny:dette:2022} and \cite{kutta:jach:kokoszka:2024}. Recently, monitoring schemes based on U-statistics have also been investigated by \cite{kirch:stoehr:2022,kirch:stoehr:2022b}. We emphasize  that all of these works focus on detecting  at most one change point of arbitrarily small size, and the procedures stop as soon as a change point has been detected.

\section{Online detection of relevant  change points} \label{sec_2}

\textbf{Structure of our approach} The procedure developed in this section has three main components. First, we propose a sequential, multiple change point method, that identifies at time $k$ in the monitoring procedure all changes that have occurred until this point. The consistency of the procedure is of independent interest and stated in Proposition \ref{lem:cp} below. 
Second, we develop a detector that identifies, among the estimated change locations from the first step, those that correspond to relevant changes—namely, those for which the mean of the corresponding segment deviates by more than $\Delta$ from the mean of the training sample. A main step to prove consistency is a weak convergence result for  the change point detector, which  can be found in Theorem  \ref{thm1}. It turns out that the limit of this statistic depends fundamentally on the entire future of the time series and in particular a standard asymptotic decision rule would require knowledge of the future before being launched. We call this procedure the \textit{oracle test}. Finally, in Section \ref{sec_22}, we present a feasible version that is based on a combination of expecting the worst (highest possible type-I error) for the future and relief (adjusting quantiles when more data is observed). In Section \ref{sec:Deltamax} we provide a data based selection method for the threshold $\Delta$ and also offer a measure-of-evidence interpretation for its magnitude. 

\subsection{An oracle test for approximate mean stability}\label{sec_21}
\textbf{Setup} Recall the change point problem and the notation  of $\Delta$-approximate mean stability  as defined by the null hypothesis in  \eqref{e:H0}, where   the piecewise constant function $\Psi$ in  \eqref{e:defpsinew}  is compared with its ``initial'' value $\Psi (1)$.
The function $\Psi$ changes at the locations
\[ 1<\theta^{(1)}<\theta^{(2)}<\cdots ,\]
where the total number of changes, say $K$,
can be $0$, finite or infinite. For ease of reference, we introduce the boundary points $\theta^{(0)}:=0,\,\, \theta^{(K+1)}:=\infty$
and set
\begin{align*} 
\Theta:=\{\theta^{(i)}: 0 \le i \le K+1\}.
\end{align*}
Then, for a stationary, centered sequence of noise variables $(\varepsilon_n)_{n \in \mathbb{N}}$, the observation at time $n$ is defined as 
\begin{align}
\label{p:defmodel}
    X_n :=  \Psi\Big( \frac{n}{N}\Big) + \varepsilon_n ,
\end{align}
where $X_1, \ldots , X_N$ correspond to the initial training sample.
\medskip

\textbf{Assumptions and notations} Throughout this paper maxima/suprema over an empty set will generically be interpreted as $-\infty$. 
For our theoretical analysis, we impose the following formal assumptions.
\begin{ass} \label{Ass1}
\item[i)] 
The time series of model errors in \eqref{p:defmodel} is centered and strictly stationary  with existing  long-run variance $\sigma^2= \sum_{n\in \mathbb{Z}} \E [\varepsilon_0\varepsilon_n] < \infty$. For some $\nu>2$ it holds that 
\[
   \sup_{n \ge k} n^{-1/\nu}\bigg|\sum_{i=1}^{n} \varepsilon_i - W(n)\bigg| = o_P(1), \qquad k \to \infty.
\] 
 Here $W$ is a centered Brownian motion with variance  $\E [W^2(1)]= \sigma^2$.
 \item[ii)] For some constant $\beta \in (0,1/2)$  there exists a $p >(1/2-\beta)^{-1}$ and $\overline{C}>0$ such that
 \[
 \mathbb{E}\Big| \sum_{i=1}^n \varepsilon_i \Big|^{p} \leq \overline{C} n^{p/2}, \qquad \forall n \in \mathbb{N}.
 \]
 \item[iii)] If $K=\infty$, the following conditions hold  for some positive constant $\underline{C}>0$
 \[
 \min_{i}|\theta^{(i)}-\theta^{(i+1)}| \ge \underline{C}, \qquad \min_i |\psi^{(i)}- \psi^{(i+1)}| \ge \underline{C}, \qquad \lim_{i\rightarrow \infty}\frac{\theta^{(i+1)}-\theta^{(i)}}{\theta^{(i)}}=0.
 \]
\end{ass}
Our assumptions can be understood in the following way: Part i) provides a Gaussian approximation for the partial sum process of errors, and assumptions of this type are standard in the literature. Such convergence results (even almost surely) follow, for example,  from KMT approximations and exist for large classes of dependent time series, such as $L^p$-$m$-approximable time series \citep[see][]{berkes:hoermann:schauer:2011} or mixing time series \citep[see][]{dehling:1983}. The second assumption is a moment bound that is standard for independent data (with sufficiently many moments) and directly extends to weakly dependent data. For $L^p$-$m$-approximability, we refer to Proposition 4 in \cite{berkes:hoermann:schauer:2011} and for mixing data to Theorem 3 in \cite{yoshihara:1978}. Finally, Condition iii) is only necessary if there exist infinitely many changes and guarantees that changes are not too close together or too small to be detected. Distances between changes can be large, but they should not grow exponentially fast (in which case $\theta^{(i+1)}-\theta^{(i)} \asymp \theta^{(i)}$). Each of these latter assumptions can be weakened to some extent, at the cost of a more involved presentation. Yet, in favor of clarity, we will not pursue further generalizations here.
\medskip

\textbf{Multiple change point detection} In order to test the null-hypothesis in \eqref{e:H0}, that is $H_0(\Delta):\sup_{1 \leq  x < \infty } |\Psi(1)-\Psi(x)|\le \Delta$,  we first identify those regions where the function $\Psi$ 
is constant and hence,
we estimate the change point locations $\theta^{(i)}$. 
For this purpose, we propose a new multiscale test statistic that compares two sums of length $h$, for different values of $h \ge 1$. The statistic has to be weighted, such  that it detects changes rapidly after they occur, all while having a small (asymptotically negligible) false detection rate. For the  sample $X_{k-2h+1}, \ldots , X_k$ of size $2h+1$ it is defined by 
\[
\gamma(h,k):= w_\beta(h,k) \cdot \Big| \sum_{n=k-2h+1}^{k-h}X_n - \sum_{n=k-h+1}^{k}X_n\Big|
\]
with weights
\begin{align} \label{e:wbeta}
w_\beta(h,k):= \frac{\sqrt{N}}{k^{1-\beta}h^\beta \log(1+k/N)}
\end{align}
where   $\beta \in (0,1/2)$  is the parameter in Assumption \ref{Ass1}ii). 
We consider data up to $\{X_1,\cdots ,X_{k}\}$ such that  $k-h \ge N$. Then,
the statistic $\gamma(h,k)$ compares the means before and after the time point $k-h$ using $h$ observations on each side. Heuristically we now want to declare that we found a new change point at the time-location $k-h$ when $\gamma(h,k)$ is suspiciously large and the multiscale change point detection procedure based on the detectors $\gamma(h,k)$ is described in Algorithm \ref{alg:change_point_estimation}. We already point out that in practice the multiple change point procedure is run at every time $k$; i.e. we scan for changes whenever a new observation becomes available.
To avoid corruption of the mean estimates, we never consider data from time points before the last detected change point. We therefore have to restrict the choices of $h$ somewhat, and to be precise we define the set of admissible indices by 
\begin{align*}
  \mathcal{S}(k, \bar k):&= \{h  \in \mathbb{N}:  \bar k \le k-2h+1\} =\Big\{1,\cdots ,\Big\lfloor \frac{k-\overline k+1}{2} \Big\rfloor \Big\} ,
\end{align*}
where $\bar k$ is the last change detected by Algorithm \ref{alg:change_point_estimation}.

\begin{algorithm}[H]
\caption{CPE}
\label{alg:change_point_estimation}
\begin{algorithmic}[1]
    \State \textbf{Input:} $\beta$ (weight parameter for \eqref{e:wbeta}), $C_{cp}$ (constant), $k$ (current time during the monitoring period)
    \State \textbf{Output:} $\widehat{\Theta}$ (estimated changes)
    
    \State Initialize $\bar k =N $ , and $\widehat{\Theta}=\{0\}$  
    
    \For{$ r= N+1, \dots, k$ }   
    \If{$ r = k$} 
        \State \textbf{Return} $\widehat{\Theta}$
    \EndIf
    \State Set $\gamma := \max_{h  \in \mathcal{S}(r,\bar k)} \gamma(h, r)$
    \If{$\gamma > C_{cp} \log(N)$}
        \State $\hat \theta \leftarrow (r-[\arg \max_{h  \in \mathcal{S}(r,\bar k)}\gamma(h, r)])/N$
        \State Update $\widehat{\Theta} \leftarrow \widehat{\Theta} \cup \{\hat \theta\}  $ 
        \State Update $\bar k \leftarrow r$.   
        
    \EndIf
    \EndFor
\end{algorithmic}

\end{algorithm}

To explain, the idea of the algorithm is to consider at time $k$ all mean-difference estimates $\gamma(h,k)$ with permissible $h$. If one of them is suspiciously large we add the associated location to the set of estimated change points, update the set $\mathcal{S}(k,\bar k)$ (i.e. update $\bar k$) of permissible indices and then continue to monitor for the next change as more data arrives. 
The output of Algorithm \ref{alg:change_point_estimation} is a set estimator     \begin{align} \label{e:defhT}
    \widehat{\Theta}(k):= CPE(\beta, C_{CP},k)=\{\hat \theta^{(1)},\hat \theta^{(2)},\cdots \}
    \end{align} 
that approximates the set of all changes that have already occurred at time $k> N$
, i.e.  
$$
\Theta(k):= \Theta \cap [0,k/N].
$$
It is helpful to notice that the set estimator $\widehat{\Theta}(k)$ is monotonically increasing in the sense that $\widehat{\Theta}(k) \subset \widehat{\Theta}(k+1)$ and  thus satisfies the relation
    \begin{align} \label{e:ThS}
        \widehat{\Theta}(k) = \widehat{\Theta}(k') \cap [0,k/N], \qquad \forall k' \ge k.
    \end{align}

We now want to formalize the consistency of the set estimator $\widehat{\Theta}(k)$ in a mathematical lemma. 
There are many ways to state such a result and we choose to formulate a uniform consistency of $ \widehat{\Theta}$ for all changes in a growing interval 
$[1, NT_N]$, where $T_N \asymp N^\kappa$. This means that for $N \to \infty$ all changes in this interval will be detected with high accuracy and the interval stands for a time period much longer than the training dataset, which includes data where  $k/N$ is in $[0,1]$. For an interpretation of the result, we need to point out two things: First, $T_N$ is not a parameter that we later need to choose. It is merely a theoretical line of demarcation between earlier and later changes. Second, the fact that our result is formulated for changes $\theta^{(i)} <T_N$ does not mean that $CPE$ ceases to detect and localize changes afterward. Consistency can in some sense also be established for later changes, though not at the same fast rate. It turns out that for technical reasons, proving rapid detection of the earlier changes is enough for our subsequent test. Throughout this paper $|{\cal A}| $  denotes the cardinality of a set ${\cal A}$.

%
%
\begin{prop} \label{lem:cp}
  {\it   Suppose that Assumption \ref{Ass1} holds
    and that $\zeta, \kappa \in (0,1/2)$ are constants such that
    \[
    \zeta>1+\kappa-\frac{1/2}{1-\beta}, \qquad  \kappa<\frac{\beta}{2(1-\beta)}.
    \]
    Moreover, let $T_N \asymp N^\kappa$ be a strictly increasing sequence satisfying $\min_i|T_N-\theta^{(i)}|>\underline{C}/2$.\\
    Then, there exists a  constant $N_0>0$ and a sequence of  events $\mathcal{E}_N$ with $\mathbb{P}(\mathcal{E}_N) \to 1$  such that the following statements hold conditionally on  $\mathcal{E}_N$ and for all $N \ge N_0$:
    \begin{itemize}
        \item[i)] For all $k=N+1,\cdots ,T_NN$ 
        \[
       |\widehat{\Theta}(k)| \le |\Theta(k)|.
        \]
        \item[ii)]         
        $
        |\Theta(T_NN)| = |\widehat{\Theta}(T_NN)|
        $
        and for the $K_N=|\Theta(T_NN)|$ elements of the respective sets, it holds that
        \[
        \max_{i=1,\cdots ,K_N}|\hat \theta^{(i)} - \theta^{(i)}| \le  N^{\zeta-1}.
        \]
        \item[iii)] 
        Whenever $N\theta^{(i)}+N^{\zeta}\leq k \leq N\theta^{(i+1)}$ for some $i$ with $\theta^{(i+1)}\leq T_N$ we have
        \[
            |\widehat{\Theta}(k)|=|\Theta(k)|.
        \]        
          \end{itemize}
          }
\end{prop}
Part i) of the lemma entails that the number of changes in the interval $[1,k/N]$ is never overestimated. Indeed, eventually all changes are detected and located with high precision, according to part ii). Notice that with the identity \eqref{e:ThS} this actually means that all changes are detected rapidly after they occur. Finally, part iii) states that if the change $\theta^{(i)}$ is sufficiently far in the past and no new change has yet occurred, the true number of changes $|\Theta(k)|$ and the number of detected changes $|\widehat{\Theta}(k)|$ are equal (with high probability).
\medskip

\textbf{An oracle test for approximate mean stability}
Up to this point, we have shown that at time $k$ in the monitoring period, we can use the CPE algorithm to get an accurate assessment of how many changes have occurred and where they are located. 
Next, we turn to the sequential estimation of the piecewise constant function $\Psi$ in \eqref{e:defpsinew}, which takes the value  
\[\Psi(x)=\psi^{(i)} \]
on the interval $(\theta^{(i-1)},\theta^{(i)}].$
Defining  the most recent change point estimate (and its scaled version) as 
\[
\hat \theta_k := \max\big[\max \{\widehat{\Theta}(k)\}, 1\big], \qquad \hat k = N\hat \theta_k~,
\]
we now define a sequential estimator at each time point $k$ by 
\begin{align} \label{e:def:psi}
\widehat \Psi(k/N)=\begin{cases}
     \frac{1}{N}\sum_{n=1}^NX_n, \quad& 1 \leq k\leq N    \\
    \frac{1}{k-\hat k}\sum_{n=\hat k+1}^{k}X_n, \quad& N<k
\end{cases}
\end{align}
so that $\widehat \Psi(k/N)$ is the average of all observed data, up to the last detected change. Recall the definition of the null hypothesis in \eqref{e:H0}, which can also be expressed as
\[   H_0(\Delta): \sup_{1 \le x <\infty}|\Psi(1)-\Psi(x)|- \Delta\le 0. \]
A natural test statistic can thus be based on the empirical version $|\widehat{\Psi}(1)-\widehat{\Psi}(k/N)|- \Delta$. We add an appropriate scaling to subsequently derive weak convergence, giving us
\begin{align} \label{e:gam}
\widehat{\Gamma}(k,\Delta):= \frac{\sqrt{N}(k-\hat k)}{k} \Big(\big|\widehat{\Psi}(1) - \widehat{\Psi}(k/N)\big|-\Delta\Big). 
\end{align}
 Naturally, if the null hypothesis $H_0(\Delta)$ is violated, then for some $k>N$ it holds that $|\Psi(1)-\Psi(k/N)|>\Delta$ and we expect that $\widehat{\Gamma}(k, \Delta)$ assumes large values. This indicates  $H_0(\Delta)$ should be rejected and to find a suitable critical values for such a decision, we study the weak convergence of the statistic 
 $$\sup_{k \ge 1}\widehat{\Gamma}(k, \Delta)
 $$ 
 in the following discussion. For the statement of the next result, we introduce the following notation.
\begin{align*} 
    \delta := \sup_{i \ge 0}|\psi^{(1)}-\psi^{(i+1)}|,\quad 
 s^{(i)}  :=  \textnormal{sign}(\psi^{(1)}-\psi^{(i+1)}),\quad
 \mathfrak{A} := \big\{i: |\psi^{(1)}-\psi^{(i+1)}|=\Delta\big\}.
    \end{align*}

\begin{theo}\label{thm1}
  {\it   Suppose that Assumption \ref{Ass1} holds and that the null hypothesis $H_0(\Delta)$, defined in \eqref{e:H0}, is satisfied. Then, there exist random variables $B,L$, such that 
    \begin{align}
    \label{pb12}
        \sup_{k \ge N+1} \widehat{\Gamma}(k,\Delta) \leq o_P(1)+ \begin{cases}
            B, \qquad \textnormal{if} \,\, \delta<\Delta,\\
            \max\{B,L\} \qquad\,\, \textnormal{if} \,\, \delta=\Delta ,           
        \end{cases}
    \end{align} 
where  
    \begin{equation}
\label{e:def:L}
\begin{split}   
    L \overset{d}{=}&\sup_{i \in \mathfrak{A}}\sup_{\theta^{(i)} \leq x \leq \theta^{(i+1)}}\frac{[s^{(i)}(x-\theta^{(i)})W_0(1)-W_i(x-\theta^{(i)})]}{x}~\\
    B\overset{d}{=}&|W_0(1)|
    \end{split}
    \end{equation} 
   and   $(W_i)_{i \in \mathbb{N}_0}$
  is a sequence   of i.i.d. copies of the Brownian motion $W$ in Assumption \ref{Ass1}(i).  \\
  Moreover, under the alternative  $H_1(\Delta)$,  that is $\delta>\Delta$, we have that 
    \begin{align*}
        \sup_{k \ge 1} \widehat{\Gamma}(k,\Delta) \rightarrow \infty.
    \end{align*}
    }
\end{theo}
$ $\\[-4ex]

Theorem \ref{thm1} motivates the following asymptotic decision rule for testing the hypotheses in \eqref{e:H0}. Suppose that $\mathfrak{A}$ is not empty and for $\alpha \in (0,1)$ denote the $(1-\alpha )$-quantile of the supremal process $\max\{B,L\} $  in \eqref{e:def:L} by $q_{1-\alpha}$. Then the decision 
\begin{align}\label{e:oracle}
    \textnormal{"reject as soon as $\widehat{\Gamma}(k,\Delta)>q_{1-\alpha}$ for some $k$"}
\end{align}
 yields a consistent, asymptotic level $\alpha$-test, in the sense that
\begin{equation*}
\begin{split}
& \limsup_{N\to \infty}  \mathbb{P}_{H_0(\Delta)}\Big(\sup_{k \ge 1}\widehat{\Gamma}(k,\Delta) >q_{1-\alpha}\Big)\le  \alpha, \\
&\liminf_{N\to \infty} \mathbb{P}_{H_1(\Delta)}\Big(\sup_{k \ge 1}\widehat{\Gamma}(k,\Delta) >q_{1-\alpha}\Big)=1.
\end{split}
\end{equation*}
In the (mathematically degenerate) case where $\mathfrak{A}=\emptyset$ a similar statement holds true for any positive number $q>0$, that is 
\begin{align*} 
\lim_N \mathbb{P}_{H_0(\Delta)}\big(\sup_{k \ge 1}\widehat{\Gamma}(k,\Delta) >q\big)= 0, \qquad \liminf_N \mathbb{P}_{H_1(\Delta)}\big(\sup_{k \ge 1}\widehat{\Gamma}(k,\Delta) >q\big)=1.
\end{align*}
Unfortunately, in practice the asymptotic quantile $q_{1-\alpha}$ is not available to the user because the distribution of $L$ depends on \textit{all change point locations $\theta^{(i)}$}. Accordingly, it cannot be known at any point in the monitoring period and hence \eqref{e:oracle} describes an oracle procedure. It will be the subject of the next section to find a feasible test procedure by constructing a surrogate for the asymptotic quantile $q_{1-\alpha}$.

\subsection{A feasible  decision rule}\label{sec_22}

\begin{algorithm}
\caption{CPE} \label{alg:feasible_test}
\begin{algorithmic}[1]
    \State \textbf{Input:} $\hat \Theta(k),  \hat{\mathfrak{A}}_N(k)$
    \State \textbf{Output:} Test Decision + Estimate of relevant CP
    
    \State Compute $\widehat{\Gamma} (k,\Delta)$ for each $k\geq N$    
    \State Compute $q_{1-\alpha}(\hat L_N(k))$
    \If{$\widehat{\Gamma} (k,\Delta)>q_{1-\alpha}(\hat L_N(k))$} 
    \State \textbf{Return} "Reject $H_0(\Delta)$."
    \State \textbf{Return}
    "Relevant change is at $\hat k$."  
    \EndIf
    
\end{algorithmic}
\caption{Testing}
\end{algorithm}

\textbf{A stochastic bound for $L$} As we have seen in the previous section, the limiting distribution $L$ is unknown and unavailable to the user at any time $k$ in the monitoring period because it depends on the entire future of the time series. Consequently we need to construct a feasible bound, using the information available at time $k$ about the hitherto observed part of the time series and some projection for the unknown future. Let us therefore define the set   
\begin{align}
\label{e:def:AN}
     \hat{\mathfrak{A}}_N(k):= &\Big\{i : |\widehat \Psi(1)-\widehat \Psi(N\hat \theta^{(i+1)})|>\Delta -\frac{\log(N)}{\sqrt{N}},\, \hat \theta^{(i+2)} \leq k/N \Big\}.
\end{align}

The set $\hat{\mathfrak{A}}_N(k)$ is an estimator of the set $\mathfrak{A}\cap [0,k]$, which contains all changes that have already occurred at time $k$ and that are equal to $\Delta$. We also define by    
\begin{align} \label{e:defk*}
   \check \theta =\max\{\Theta(k)\}, \quad \check k=N\check \theta        
\end{align} 
 the last  scaled and unscaled change point that has occurred.  With these notations we consider the statistics
\begin{align}
&  \hat L_{N,1}(k) =\sup_{i \in \hat{\mathfrak{A}}_N(k)}\sup_{\hat \theta^{(i)} \leq x \leq \hat \theta^{(i+1)}}\frac{\sqrt{N}}{Nx}\hat s^{(i)}\Big((x-\hat \theta^{(i)})W(N)-\Big[W(Nx)-W(N\hat \theta^{(i)})\Big]\Big)\\
    & \hat L_{N,2}(k)= \sup_{x \geq \hat \theta_k}\sup_{ \hat \theta_k\leq \theta \leq x} \frac{\sqrt{N}}{Nx}\Bigg|(x-\theta)W(N)-\Big[W(Nx)-W(N\theta)\Big]\Bigg|,
    \label{hd1}
\end{align}
where $W$ is the Brownian motion from Assumption \ref{Ass1} and
\begin{align*} 
 \hat s^{(i)} =\text{sign}(\widehat \Psi(1)-\widehat \Psi(N\hat \theta^{(i+1)})).
\end{align*}
The desired upper bound is then given by
\begin{align}     \label{e:def:Lh}
   \hat L_N(k)=\max\{\hat L_{N,1}(k),\hat L_{N,2}(k)\}~.
\end{align}

Let us give a more detailed motivation of this construction. For each $k$ we divide $L$ into two parts, i.e.
\begin{align}
    L=\max\{L_1(k),L_2(k)\}
\end{align}
where \begin{align}
     L_1(k)=\sup_{\substack{i \in \mathfrak{A}\\N\theta^{(i+1)}\leq k}}\sup_{\theta^{(i)}\leq x \leq \theta^{(i+1)}}\frac{[s^{(i)}(x-\theta^{(i)}W_0(1)-W_i(x-\theta^{(i)})]}{x}\\
    L_2(k)=\sup_{\substack{i \in \mathfrak{A}\\N\theta^{(i+1)}> k}}\sup_{\theta^{(i)}\leq x \leq \theta^{(i+1)}}\frac{[s^{(i)}(x-\theta^{(i)}W_0(1)-W_i(x-\theta^{(i)})]}{x}
\end{align}
and note that $\hat L_{N,1}(k) $ approximates the distribution of the random variable $L_1(k)$  over the already observed (rescaled) time span $[1,k/N]$, while  $\hat L_{N,2}(k)$  provides  an upper bound for $L_2(k)$ for the unobserved period $(k/N,\infty)$. Notice that in $\hat L_{N,2}(k)$, we take a  supremum with respect to  $x$ (the most extreme time point) and a supremum with respect to  $\theta$ (the most adverse position of a change). Combining these two variables at time $k$ yields $\hat L_N(k)$, a feasible distribution (see the discussion preceding equation \eqref{pb30}) that is stochastically dominating the distribution of $L$, except for a negligible error.
\begin{prop}
\label{l2}
 {\it    Suppose that Assumptions \ref{Ass1} hold. 
    Then, under $H_0(\Delta)$, we have 
    \begin{align}
    \label{pb11}
    \limsup_{N \to \infty}  \mathbb{P} \Big (\sup_{k >N}\hat L_N(k) \le  t  \Big )  \le \mathbb{P}\big(\max(B,L) \le t\big) ,\qquad \textnormal{for all} \quad t \in \mathbb{R}.
    \end{align}    
    }
\end{prop}
\begin{rem} $ $
     The proof of Proposition  \ref{l2} is based on an explicit coupling of the random variables $\hat L_N(k)$ and $L$ on a suitable probability space. More precisely, on this space, we construct a third (infeasible) variable $ \hat L_N  $  with the same distribution as 
        \begin{align}    \label{e:def:coupling}
  \sup_{i \in \hat{\mathfrak{A}}_N(NT_N)}\sup_{\hat \theta^{(i)} \leq x \leq \hat \theta^{(i+1)}}\frac{\sqrt{N}}{Nx}\hat s^{(i)}\Big((x-\hat \theta^{(i)})W(N)-\Big[W(Nx)-W(N\hat \theta^{(i)})\Big]\Big)~,
\end{align} 
where $T_N$ is the constant from Proposition  \ref{lem:cp}. We then show that 
\[
\hat L_N(k) \ge \hat L_N \ge L+o_P(1),
\]
allowing us to make the dominated probability statements in Proposition   \ref{l2} for any $k$.\smallskip
\end{rem}
\textbf{Estimating the long-run variance and feasibility of the distribution} The distribution of  $\hat L_N(k)$ depends, with the exception of the long-run variance  $\sigma^2$ of $(\varepsilon_i)_{i \in \mathbb{N}}$, only on quantities available at time $k$. Using the training data $X_1,\cdots ,X_N$ it is straightforward to define an estimator for $\sigma^2$. For example, using the blocking technique from  \cite{Wu2007}, we define with the sums \(S_{j,k} = \sum_{i=j}^k X_{i,n}\) the long-run variance estimator
\[
\hat{\sigma}^2 := \frac{1}{\lfloor N / m_N \rfloor - 1} \sum_{j=1}^{\lfloor N / m_N \rfloor - 1} \frac{\left(S_{(j-1)m_N+1,jm_N} - S_{jm_N+1,(j+1)m_N}\right)^2}{2m_N}.
\]
Here \(m_N\) denotes a divergent sequence of integers that is chosen proportional to  \(N^{1/3}\) for theoretical reasons. Under our assumptions is follows that (see Lemma 5.6 in \cite{bastian:dette:2025})
\[
\label{pb30}
\hat{\sigma}^2 = \sigma^2 + O_P(N^{-1/3}).
\]
Using this estimator then allows us to calculate a good approximation of the quantiles  $q_{1-\alpha}^{(N,k)}$ of $\hat L_N(k)$. To be precise we define
\begin{align}
    \hat q_{1-\alpha}^{(N,k)}=\hat \sigma\Big[\frac{1}{\sigma} q_{1-\alpha}^{(N,k)}\Big]
\end{align}
and note that $\frac{1}{\sigma}q_{1-\alpha}^{(N,k)}$ are the quantiles of the pivotal process $\hat L_N(k)/\sigma$. 
 In particular, we can calculate them with arbitrary precision by Monte Carlo methods.  Having furnished an appropriate estimator for $\sigma^2$ hence grants us access to (a good approximation of) the quantiles of $\hat L_N(k)$. 
\begin{lemma}
    \label{lem:var}
 {\it     Suppose that Assumption \eqref{Ass1} holds. Then it follows that
    \begin{align}
        \max_{k\geq 1}\Big|\hat q_{1-\alpha}^{(N,k)}-q_{1-\alpha}^{(N,k)}\Big|=o_P(1).
    \end{align} 
    }
\end{lemma}
\textbf{A bootstrap test decision rule} We now apply Proposition  \ref{l2} to define a feasible decision rule for the  hypotheses in \eqref{e:H0}. First notice that for any $N$ the feasible random variables $\hat L_N(k)$ are decreasing  in $k$. In particular, it follows that   $\hat q_{1-\alpha}^{(N,k)} \ge  \hat q_{1-\alpha}^{(N,h)}\ge q_{1-\alpha}+o_P(1)$ for $k \le h$ (with $q_{1-\alpha}$ being the upper $\alpha$ quantile of the random variable $L$ defined in \eqref{e:def:L}). The next corollary gives a large sample justification for using $q_{1-\alpha}^{(N,k)}$ in the test decision.
\begin{cor}
    \label{lem:var}
   \textit{ Suppose that Assumption \eqref{Ass1} holds. Then it follows that
    the decision rule 
    \begin{align}\label{e:bootsor}
        \textnormal{"reject as soon as $\widehat{\Gamma}(k,\Delta)>\hat q^{(N,k)}_{1-\alpha}$ for some $k$"}
    \end{align}
     yields a consistent, asymptotic level $\alpha$-test, in the sense that
    \begin{align*} 
    \limsup_{N \to \infty}  \mathbb{P}_{H_0(\Delta)}\Big(\sup_{k \ge 1}\widehat{\Gamma}(k,\Delta) >\hat q^{(N,k)}_{1-\alpha}\Big)\le  \alpha, \qquad \liminf_{N \to \infty} \mathbb{P}_{H_1(\Delta)}\Big(\sup_{k \ge 1}\widehat{\Gamma}(k,\Delta) >\hat q^{(N,k)}_{1-\alpha}\Big)=1.
    \end{align*}
    }
\end{cor}
\smallskip
A lingering concern might be the computational budget needed to update the quantiles $\hat q_{1-\alpha}^{(N,k)}$ for each new incoming data point. Closer inspection, however, reveals that $\hat L_N(k)$ in fact only changes when a new change point is detected at time $k$. Consequently,  updating the quantiles is required at most once for every $\underline CN$ observations (under Assumption \ref{Ass1}) which does not impose an undue computational burden.

\subsection{Choosing $\Delta$ and monitoring the deviation from mean stationarity} \label{sec:Deltamax}
The choice of $\Delta$ plays an important role in the methodology we have proposed. In most contexts, subject-specific knowledge will yield reasonable choices for $\Delta$ (see our example in the introduction). However, such knowledge is not always available and data-based choices of $\Delta$ may be desirable. In the following we provide a discussion and some associated extensions of our methodology that are pertinent to applying it. \medskip

In many applications continued monitoring even after detecting a relevant change might be warranted. An extensive example of this will be discussed in the next section. In such cases considering multiple $\Delta$ is of interest, i.e. to consider the hypotheses \eqref{e:H0} for multiple $\Delta$. Naively implemented, this would entail  multiple-testing penalties. However, since the hypotheses of interest are nested ($H_0(\Delta) \subset H_0(\Delta')$ for $\Delta<\Delta'$) these penalties can be avoided by some minor adjustments of our procedure, where we make the test decisions monotone (rejection of $H_0(\Delta)$ implies rejection of $H_0(\Delta')$). For this purpose we must rid the  quantiles $\hat q_{1-\alpha}^{(N,k)}$  off their dependence on the specific threshold $\Delta$. Such a modification is detailed below.

Note that the dependence of $\hat q_{1-\alpha}^{(N,k)}$ on $\Delta$ stems from the definition of $\hat{\mathfrak{A}}_N(k)$. We may instead use the estimator 
\begin{align}
    \tilde{\mathfrak{A}}_N(k):=& \Big\{i: |\hat \Psi(1)-\hat \Psi(N\hat \theta^{(i+1)})|\\
    &\quad\quad\geq \max_{l:\hat \theta^{(l+2)} \leq k/N} |\hat\Psi(1)-\hat \Psi(N\hat \theta^{(l+1)})|-\frac{\log(N)}{\sqrt{N}}, \hat \theta^{(i+2)}\leq k/N\Big\} \nonumber
\end{align}
in the definition of $\hat L_N(k)$ and its associated quantities which thereby do not depend on $\Delta$ anymore. A detailed motivation for the definition of $\tilde{\mathfrak{A}}_N(k)$ is given in Appendix \ref{sec:motiv1}.
It is straightforward to establish that similar guarantees as those in Proposition   \ref{l2} and the subsequent discussions hold for this version of $\hat L_N(k)$ and its quantiles, which we shall denote by  $\tilde q_{1-\alpha}^{(N,k)}$. We now observe the following three facts:
\begin{itemize}
    \item[(1)] $\widehat{\Gamma}(k,\Delta)$ is monotone in $\Delta$.
    \item[(2)] The quantiles $\tilde q_{1-\alpha}^{(N,k)}$ do not depend on $\Delta$.
    \item[(3)] The hypotheses $H_0(\Delta)$ are nested.
\end{itemize}
The  sequential rejection principle therefore yields that we may test for multiple $\Delta$ simultaneously without incurring multiple-testing penalties. But much more is true; the sequential nature of our procedure and its theoretical guarantees in fact imply that, for each time point $k$, we may find a maximal
\begin{align}
    \widehat{\Delta}_{\max} (k)=\sup \big \{ \Delta\geq 0 ~\big | ~\widehat{\Gamma}(k,\Delta)>\tilde q_{1-\alpha}^{(N,k)}\big \} ~, \label{delta_maxK}
\end{align}
for which 
\begin{align}
\label{LocHyp}
    H_0(\Delta,k)=\sup_{\check \theta\leq x \leq k/N}|\Psi(1)-\Psi(x)|\leq \Delta
\end{align}
is rejected at significance level $\alpha$. In this sense $\widehat{\Delta}_{\max} (k)$ serves as a \textit{sequential measure of maximal deviation from mean stationarity at time $k$.}  \\
\begin{rem} $ $
\begin{enumerate}
    \item[(1)] Rejecting the local hypothesis \eqref{LocHyp} of course also implies rejecting the global hypothesis \eqref{e:H0}. Yet, in many applications (see our data example below) it might be desirable to continue monitoring even after detecting a first relevant change. In such cases knowing whether or not the local hypothesis $H_0(\Delta,k)$ is rejected may be of greater interest. 
    \item[(2)] More generally one might be interested in simply monitoring $\widehat{\Delta}_{\max} (k)$ to evaluate how far the current mean deviates from its historical benchmark, for example in the applications in \ref{sec_32} and \ref{App: diabetes example}. Extensions to other benchmarks such as the mean on some other historical segment might also be of interest but are not further investigated here.
\end{enumerate}
\end{rem}

\section{Finite sample properties} \label{sec_3}
In this section, we investigate the finite sample properties of the proposed methodology for monitoring an incoming stream of data for changes larger than a relevant size $\Delta.$ Our aim is not to present comprehensive numerical experiments. Rather, we want to demonstrate that the new theory developed in this paper is likely to have practical significance.

\subsection{Simulation study} 

\textbf{Setup} We will generate data with training sample sizes $N=50,100,200$. Monitoring is in theory a never-ending process, but for our simulations a finite time horizon has to be chosen, which we fix at $n=20N$; much larger than the initial training dataset. For our simulations we need to specify a set of change point locations $\Theta$ and their size. To investigate a range of different setups, we draw the change points randomly  in every simulation run as follows:
\begin{enumerate}
    \item[(1)] The number of changes $l$ is chosen randomly according to $l \sim \text{Unif}(2,\cdots ,6)$
    \item[(2)] The change locations $N\theta^{(i)}$ are chosen uniformly from $N<k<19N$ under the constraint
    \begin{align*}
        N(\theta^{(i+1)}-\theta^{(i)})&\geq 90.       
    \end{align*} 
    More precisely, we uniformly sample $l$ locations, trying repeatedly until the obtained sample satisfies the above constraint.
    \item [(3)] In the initial training phase we set the population mean to $\Psi(0)=0$. Afterwards, we draw the differences $\delta_i:=\Psi(\theta^{(i)})-\Psi(1)$ randomly as specified below and notice that the function $\Psi$ is uniquely characterized by the $\Psi(1), (\delta_i)_i$ and the change point locations. Next, we generate the values for $\delta_i$ at random. In view of the distinction by cases made in Theorem \ref{thm1} there exist three  cases, depending on $\delta = \max_i |\delta_i|$ being smaller, equal or larger than $\Delta$. 
        \begin{itemize}
            \item \textbf{Null Hypothesis Interior} ($\delta<\Delta$):\\ $\delta_i \sim \text{Uniform}(0.1, \, \Delta - 0.1). $            
            \item \textbf{Null Hypothesis Boundary} ($\delta=\Delta$):\\ $\delta_i = \Delta \cdot \mathcal S$, where $\mathcal{S} \sim \text{Uniform} (\{-1, 1\})$ .
            \item \textbf{Alternatives} ($\delta>\Delta$):\\ 
            Under the alternative, some, but not all break sizes $|\delta_i|$ have to be larger than $\Delta$. For this reason, we flip a coin for each change whether it is bigger or smaller than $\Delta$ (under the constraint that at least one of them is larger than $\Delta$).  
            The magnitudes of the larger changes are then drawn as follows (where Alt I.-Alt III. represent different scenarios for the alternative that are investigated separately) 
            \begin{align}
                \delta_i \sim \text{Uniform}\begin{cases}
                    (\Delta + 0.1, \, \Delta +1.0) &\text{~~~~Alternative I}\\
                    (\Delta+0.5 , \Delta+1.5) &\text{~~~~Alternative II}\\
                    (\Delta+1 , \Delta+2) &\text{~~~~Alternative III}.
                \end{cases}
            \end{align}
            The magnitudes of the smaller changes are drawn as follows 
            \begin{align}
              \delta_i \sim \text{Uniform}(0.1, \, \Delta - 0.1).
            \end{align}
        \end{itemize}    
\end{enumerate}
For the error process $(\varepsilon_i)_{i \in \N}$ we consider three different possibilities given by
\begin{align}
\label{IID}(\text{IID}) \quad &~ \varepsilon_i= \tfrac{1}{2}\eta_i\\
\label{MA}  (\text{MA)} \quad &~ \varepsilon_i = \tfrac{1}{\sqrt{5}}\big( \eta_i + \tfrac{1}{2}\eta_{i-1}\big)\\
\label{AR} (\text{AR})  \quad &~ \varepsilon_i = \tfrac{\sqrt{3}}{4} \big( \eta_i + \tfrac{1}{2}\varepsilon_{i-1}\big),
\end{align}
where $(\eta_i)_{i\in\mathbb{Z}}$ is an i.i.d.\ sequence of standard normal random variables. Notice that in all scenarios normalization factors are chosen such that $\text{Var}(\varepsilon_i)=1/4$ to make results comparable. 
\medskip

\textbf{Results} We record the empirical rejection rates in Table \ref{tab:combined_results} below. For the parameter $\beta $ in Algorithm \ref{alg:change_point_estimation} we consider the choices $\beta \in \{0.1,0.3, 0.45\}$ and we choose a nominal level of $\alpha = 5\%$ in all cases under consideration. Each update of the quantiles is based on $100$ bootstrap replications with a time horizon of $x \leq 20$ for the calculation of $\hat L_{N,2}$. All rejection rates are based on $600$ simulated datasets and the  threshold   $\Delta=1$.

\renewcommand{\arraystretch}{1.3}
\begin{table}[H]
\centering
\small
\setlength{\tabcolsep}{8pt} 
\begin{tabular}{lcccccccccc}
\toprule
& & \multicolumn{3}{c}{$\beta = 0.1$} & \multicolumn{3}{c}{$\beta = 0.3$}& \multicolumn{3}{c}{$\beta = 0.45$} \\
\cmidrule(lr){3-5} \cmidrule(lr){6-8} \cmidrule(lr){9-11}
& $(N)$ & $50$ & $100$ & $200$ & $50$ & $100$ & $200$ & $50$ & $100$ & $200$ \\
\midrule
\multirow{5}{*}{\textbf{IID}} 
& Interior            & 0.00 & 0.00 & 0.00 & 0.00 & 0.00 & 0.00 &0.00 &0.00 &0.00\\
& Boundary        & 0.03 & 0.02 & 0.04 & 0.03 & 0.01 & 0.02 &0.02 &0.03 & 0.05\\
& Alternative I   & 0.54 & 0.57 & 0.69 & 0.65 & 0.64 & 0.85 & 0.72&0.79 &0.88\\
& Alternative II  & 0.78 & 0.87 & 0.93 & 0.84 & 0.88 & 0.94  & 0.85 &0.91 & 0.94\\
& Alternative III & 0.92 & 0.96 & 0.96 & 0.96 & 0.97 & 0.98 & 0.96 & 0.97 &0.97  \\
\midrule
\multirow{5}{*}{\textbf{AR(1)}} 
& Interior            & 0.00 & 0.00 & 0.00 & 0.00 & 0.00 & 0.00 & 0.00 &0.00 &0.00\\
& Boundary        & 0.05 & 0.04 & 0.04 & 0.04 & 0.03 & 0.03 & 0.10 & 0.02 &0.01\\
& Alternative I   & 0.34 & 0.43 & 0.52 & 0.43 & 0.56 & 0.71 &0.72 &0.8 &0.87\\
& Alternative II  & 0.71 & 0.84 & 0.92 & 0.74 & 0.89 & 0.95 &0.79 &0.94&0.98\\
& Alternative III & 0.82 & 0.92 & 0.96 & 0.83 & 0.94 & 0.98 & 0.86 & 0.98 & 0.99\\
\midrule
\multirow{5}{*}{\textbf{MA(1)}} 
& Interior            & 0.00 & 0.00 & 0.00 & 0.00 & 0.00 & 0.00 &0.00 &0.00 &0.00\\
& Boundary        & 0.00 & 0.00 & 0.01 & 0.04 & 0.03 & 0.02 & 0.10 & 0.04  & 0.03\\
& Alternative I   & 0.54 & 0.57 & 0.67 & 0.69 & 0.66 & 0.89 & 0.70 & 0.83 &0.87 \\
& Alternative II  & 0.69 & 0.86 & 0.89 & 0.77 & 0.90 & 0.95 &0.81 &0.94&0.97\\
& Alternative III & 0.82 & 0.92 & 0.96 & 0.84 & 0.94 & 0.98 & 0.80 & 0.92 & 0.94 \\
\bottomrule
\end{tabular}
\caption{\it Empirical rejection rates of the test \eqref{e:bootsor} for the scenarios described at the beginning of the section. 
}
\label{tab:combined_results}
\end{table}

We observe that the empirical sizes never exceed the nominal level of $5\%$ except for the combination of very low sample size ($N=50$) and $\beta$ very close to 0.5. In the interior of the null hypothesis the type-I error is always $0$, which is a desirable property and can be explained by   the asymptotic degeneration of the test statistic according to Theorem \ref{thm1}. On the boundary 
we see an acceptable approximation of the nominal level in most cases (typically from below), indicating that the bound \eqref{pb11} is almost tight. In the alternative settings we observe that rejection rates are reasonable even for low sample sizes and that they increase as the sample size and the mean differences increase. In view of the complicated composite hypotheses and the fact that the asymptotic distribution relies on the unknown future of the monitored process, we observe overall a surprisingly good empirical performance for all settings under consideration.




\subsection{Data analysis I: political polls in Germany} \label{sec_32}

\textbf{Political polling as real-time monitoring} We consider the problem of monitoring polling averages for political parties over time. In many democracies, political polls are available almost continuously (the analyzed datasets contain polls appearing at least every 2-3 days) and are constantly used by party analysts to modify and refine political messaging - especially when approaching state and local elections. Because  polls are inherently random, most analysts are careful not to over-interpret the result of any single poll, but rather bundle and interpret recent polls collectively. This can be seen as a kind of informal local averaging. The methods presented in this paper can be seen as a related but systematized approach, granting statistical guarantees for the performed inference.

The fluctuations across polls can reasonably be modeled as independent, using simple random sampling as an approximation of reality. The polling average and, more fundamentally, voter preferences evolve dynamically over time. In this work they correspond to the mean parameters in our statistical model \eqref{e:defpsi}. The aim of the political analyst is to perceive substantial changes in voter preferences in real-time to react to them accordingly. \textit{Since voter preferences are never constant over time, it is not reasonable to test for a perfectly constant mean}, but rather for deviations that are large enough to merit political reaction. The definition of what constitutes a "relevant deviation" in polling averages will differ based on the situation. In our analysis here, we focus on evaluating the evolution of 
the sequential measure $\widehat{\Delta}_{\max} (k)$ of deviation from mean stationarity at time $k$
over time. In terms of \eqref{delta_maxK} this is the smallest value of   $\Delta$, for which we can reject the null hypothesis at time 
 $k$ based on the evidence since the last change, yielding a local measure of deviation from the historical voting share. For a detailed description see Section \ref{sec:Deltamax}.\\

\textbf{German polling data} 
The following analysis is based on the Infratest-dimap Election Polling Data\footnote{The data can be accessed at \url{https://dawum.de/Bundestag/Infratest_dimap/2025-02-06/\#Umfrageverlauf}.}, which provides a dense historical  time series of polling results in Germany. The data combines surveys conducted by various polling institutes, tracking party support over time. All dates are in the format dd.mm.yyyy.
New surveys are typically published every other day, and over the time frame analyzed in this paper, from 01.04.2020 to 01.01.2023, 625 polls are available in total. Of course we cannot truly perform a real-time analysis on a historical dataset. What we can do, however, is to conduct a \textit{counterfactual study, which demonstrates how our methods would have performed if they had been applied in the past.} As we will see below, our methods perform well in rapidly detecting changes and distinguishing small ones from truly relevant ones. In contrast, standard sequential tests are unsuitable for this task: In a dynamic environment, such as with polling data, they inevitably detect some small non-stationarity, typically right after launching and stop monitoring (see Table \ref{tab:my_label} for the time the respective procedures stop). Accordingly, they miss out on the more substantial, meaningful changes that occur later during the monitoring period.

Since Germany has a multi-party system, we confine our analysis to two parties represented in parliament: the Christian Democrats (union of CDU/CSU)  and the Green party (die Grünen).  \\
\begin{table}[h]
    \centering
    \begin{tabular}{|c|c|c|c|}
    \hline
       Method  &  Greens (\# Obs.)  & CDU/CSU (\# Obs.)  \\
       \hline\hline
        HKHS & 04-08-2020 $(k=1)$ & 28-08-2020 $(k=15)$\\
        KD & 25-09-2020 $(k=30)$ &29-08-2020 $(k=16)$\\
        DG & 29-09-2020 $(k=33)$& 13-09-2020 $(k=24)$\\
        F &  03-10-2020 $(k=35)$& 13-09-2020 $(k=24)$\\
        \hline
    \end{tabular}
    \caption{\it Detected  change points  by four  comparative methods developed by \cite{horvath:kokoszka:huskova:steinebach:2003} (HKHS), \cite{fremdt:2015} (F), \cite{dette:gossmann:2019} (DG), and \cite{kutta:doernemann:2025} (KD). 
  In the parentheses, we denote the number of observations (\# Obs.) after the start of monitoring at which the change was detected.  
  } 
    \label{tab:my_label}
\end{table}

\begin{figure}[ht]
    \centering
    \includegraphics[width= \linewidth]{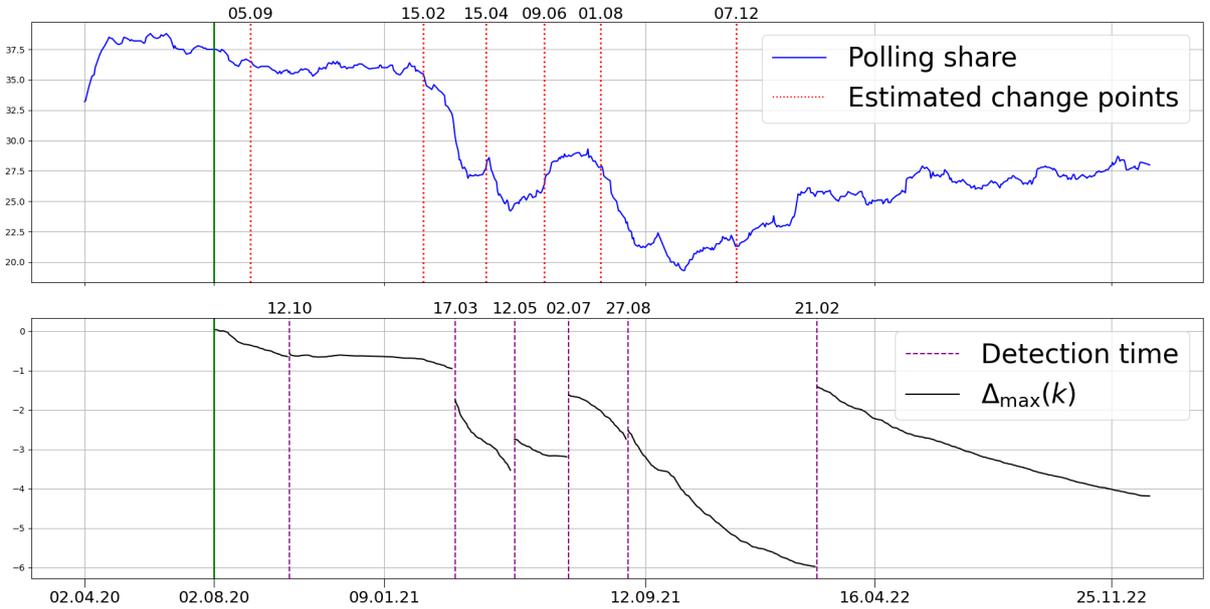}   
    \caption{\it  Polling share for the CDU/CSU fraction. The upper graph contains the time series and the change point location estimates (with corresponding dates on top).  The lower graph contains $\widehat{\Delta}_{\max} (k)$ multiplied by the sign of the detected deviation from the initial mean. Additionally it also contains the times where the changes were located (similarly with the dates of detection on top). The green line marks the beginning of the monitoring period. Total sample size is $625$ with $N= 62$ for the benchmark period.}
     \label{fig:CDU}
\end{figure}

\begin{figure}[t]
    \centering
    \includegraphics[width=\linewidth]{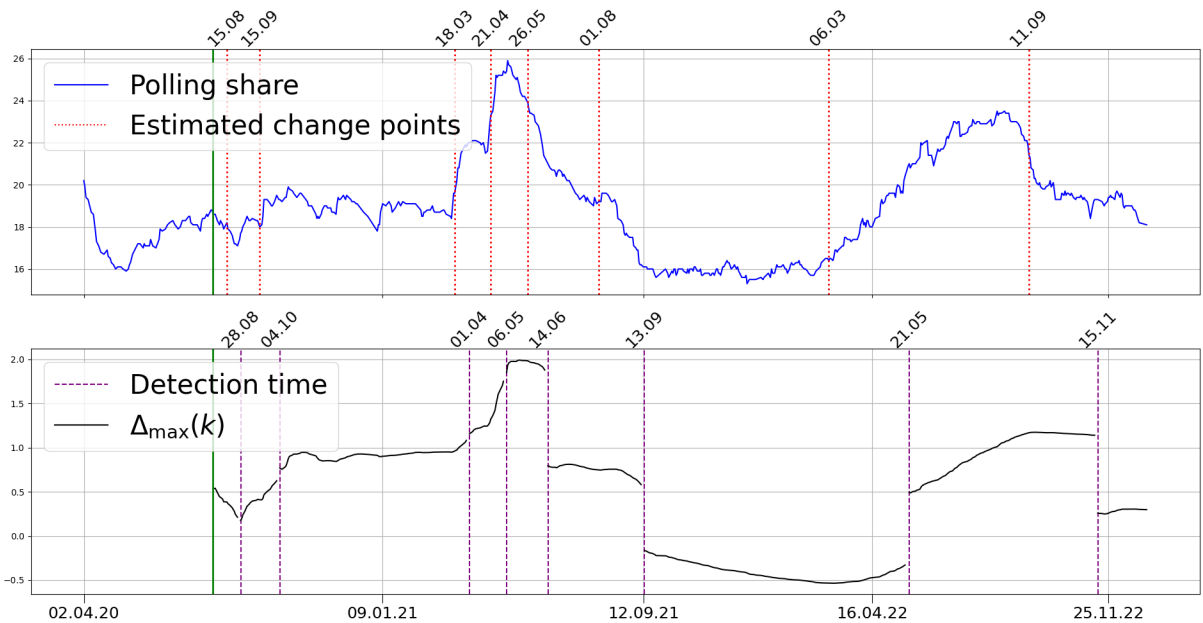}
    \caption{\it Polling share for the green party. The upper graph contains the time series with estimates of the change point locations (corresponding dates on top). The lower graph contains $\widehat\Delta_{\max} (k)$ multiplied by the sign of the detected deviation from the initial mean. Additionally, the vertical dashed lines denote the times where the changes were detected (similarly with dates on top). The green line marks the beginning of the monitoring period. As before, the total sample size is 625 and $N = 62 $ for the benchmarking period.  }
    \label{fig:Greens}
\end{figure}

For our procedure, we need to choose the parameter $\beta$ as in \eqref{e:wbeta} that is required for  Algorithm \ref{alg:change_point_estimation}. As we may reasonably assume that our observations are independent and bounded it is clear that Assumption (ii) holds for any $p\geq 1$ and we therefore may choose any $\beta \in (0,1/2)$. Therefore, we choose $\beta=0.45$ to minimize the detection delay. We begin with the CDU/CSU  polling data which is displayed in Figure \ref{fig:CDU}, followed by the Green Party in Figure \ref{fig:Greens}. 
These  datasets contain a total of $625$ observations with an initial sample of   $10\%$, that is $N=62$,  of them.
In the top plot of both Figure \ref{fig:CDU} and \ref{fig:Greens}, the observed data is shown (blue curve) together with the locations of the estimated change points by our procedure, which are demarcated by the  vertical dashed red lines.  The beginning of the monitoring period is marked by the bold green line.  In the bottom plot of this figure, we  display the time points when the changes are  detected by  dashed vertical lines (in purple). Note that to every purple line (the first time when a new change is detected) there corresponds a red line (estimate of the  location of the change point). The red line is dated earlier than the purple line because we can only detect changes with a delay. With the black curve in the bottom part of Figure \ref{fig:CDU} we display the detected deviation   $\widehat{\Delta}_{\max} (k)$ from mean stationarity (for different $k$), as defined in  \eqref{delta_maxK}, multiplied by the sign of the detected deviation. As 
$\widehat{\Gamma}(k,\Delta)$ only incorporates data that occurs after the last detected change  point  we generally expect $\widehat{\Delta}_{\max} (k)$ to need some time to adjust to its true value.
\smallskip


We stress that the following discussion is not a political analysis; it only serves to showcase the capabilities of the proposed method on a real-world dataset. To discuss the practical value of our method we associate the observed change points as well as the behavior of $\widehat{\Delta}_{\max} (k)$ to contemporaneous events that were covered in the German news.


\textbf{CDU/CSU:} Before our analysis, we notice that in the observed time frame, the CDU/CSU starts from a local polling high and thus all subsequent changes have a negative sign compared to the initial period. 
The monitoring begins with a relatively stable period where $\widehat{\Delta}_{\max} (k)\approx 0.75\%$ until the second change is detected on the 22.03.2021 with a delay of about one month (i.e. dated to the 15.02.2021).  Here $\widehat{\Delta}_{\max} (k)$ leaps to $\approx$ 2\% and then increases quickly to almost 4\% as more data is collected. This change appears  concurrently with the decisions to prolong the Covid-19 lockdown measures on the 10.02.2021 as well as the public corruption allegations against the CSU politician Georg Nüßlein on the $25^{th}$ February. During this time period, the CDU/CSU saw plummeting polling averages. 
For the next months $\widehat{\Delta}_{\max} (k)$ retains a relatively stable value of about 3\%. Nonetheless one further change, dated to the 15.04.2021, is detected on the 10.05.2021. The fact that $\widehat{\Delta}_{\max} (k)$ stays relatively stable despite this new change indicates that the deviation has actually increased (as even a small amount of data indicates a large $\widehat{\Delta}_{\max} (k)$). The change on the 10.05.2021 coincides quite closely with the announcements of Armin Laschet and Markus Söder to be available as chancellor candidates in the upcoming national elections. The next change is detected on the 01.07.2021 and dated to the 09.06.2021. We observe that $\widehat{\Delta}_{\max} (k)$ decreases to about 1.5\% amid an uptick in polling results for the Christian Democrats. 
The date, 09.06.2021, coincides closely with the state election in Saxony-Anhalt which was won by the CDU with a result of 37\%, gaining 7 percentage points over their previous results in the state. The next change is detected on the 26.08.2021 and dated to the 01.08.2021, but $\widehat{\Delta}_{\max} (k)$ starts sharply decreasing already sometime in July. Around that time (17.07.2021) Armin Laschet, then-chancellor candidate of the CDU/CSU, was filmed laughing in the background of a speech in Ahrtal that was recently devastated by floods. The event generated an immense (negative) media response and a downward trend in polling results. This trend is reflected by an increasing $\widehat{\Delta}_{\max} (k)$ and continues until the 22.03.2022 with its curve starting to flatten somewhere around the end of December 2021/beginning of January 2022 until a new change is picked up, dated to the 07.12.2021 which coincides with the (first ever) member survey of the CDU/CSU from the 04.12 to the 16.12. As a consequence $\widehat{\Delta}_{\max} (k)$ reduces from 6\% to 1.5\%. It slowly increases again as more data is collected, leveling off at about 4\% which is reflected by the relatively constant polling results of about 25-27\%.

\begin{figure}
    \centering
    \includegraphics[width=0.75\linewidth]{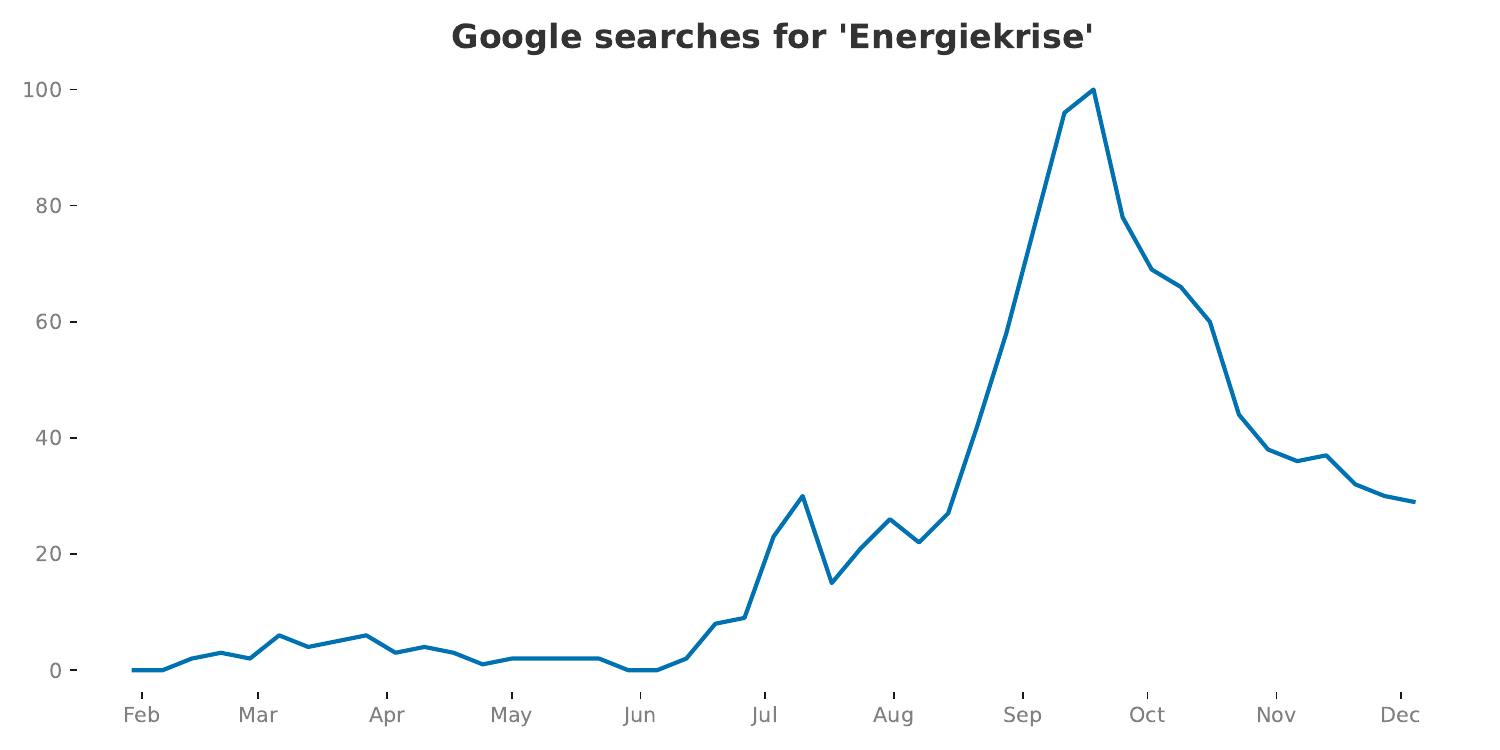}
    \caption{ Relative search frequency of the term "Energiekrise" (energy crisis) on Google during 2022. Data source: Google Trends (https://www.google.com/trends)}
    \label{fig:crisis}
\end{figure}

\textbf{Green Party:} Owing to the relatively unstable benchmark period of the dataset, the procedure detects two changes at the beginning of the monitoring period. However, $\widehat{\Delta}_{\text{max}} (k)$ does not change too drastically- it varies between 0.5\% and 1\%. Starting in late March 2021, we observe a sudden increase to about 2\% that is accompanied by two change points which are detected on the 31.03.2021 and 05.05.2021 and dated to the 18.03.2021 and 21.04.2021, respectively. The first change point likely corresponds to the state elections in Baden-Württemberg and Rheinland-Pfalz on the 14.03.2021, where the Greens gained 2 percentage points and 4 percentage points compared to the last election cycle, respectively. The second change point coincides with the election of Annalena Baerbock as the Greens chancellor candidate. The next change is detected on the 12.06.2021 and dated to the 26.05.2021; we also observe that $\widehat{\Delta}_{\text{max}}(k)$ suddenly stops increasing and levels off at 2\% in May. In this time frame, media outlets started reporting about allegations regarding misleading information in her curriculum vitae and undeclared bonus payments. $\widehat{\Delta}_{\text{max}}(k)$ remains at a stable 0.75\% during this period until a further change is detected on the 12.09.2021 and dated to the 01.08.2021. $\widehat{\Delta}_{\text{max}}(k)$ drops to 0 upon detection of the change and increases as more data is incoming until it reaches 0.5\%. Note however that the sign of the mean difference is flipped during this period, which is also reflected in the lower polling results during this period. Several possibly related events take place at the beginning of August, such as the election disqualification of the Green party in the state Saarland and a concerted anti-Greens advertisement campaign by the Conservare Communications GmbH. $\widehat{\Delta}_{\text{max}}(k)$ then increases slowly from March 2022 until May 2022 where a new change is detected on the 20.05.2022 and dated to the 06.03.2022, which also prompts $\widehat{\Delta}_{\text{max}}(k)$ to jump up to 0.5\%. This increase could potentially be attributed to the increased media presence of Annalena Baerbock in her function as minister of foreign affairs due to the Russian invasion of Ukraine in late February 2022. $\widehat{\Delta}_{\text{max}}(k)$ continues to increase until August 2022 where it then plateaus until a final change point is found on the 13.11.2022 that is dated to the 11.09.2022. The beginning of the plateau and the time of the change point coincide with an increased public interest in the German energy crisis (see Figure~\ref{fig:crisis} where Google search trends for the term ``Energiekrise'' are displayed). The crisis consisted of spiked energy prices and anticipation of potential energy shortages. In the media it was often associated with the ministry of economic affairs and its minister Robert Habeck (Green party).

 \subsection{Data analysis II: diabetes dataset continued}
\label{App: diabetes example}
\begin{figure}[h]
\centering
\includegraphics[width = 1\linewidth]{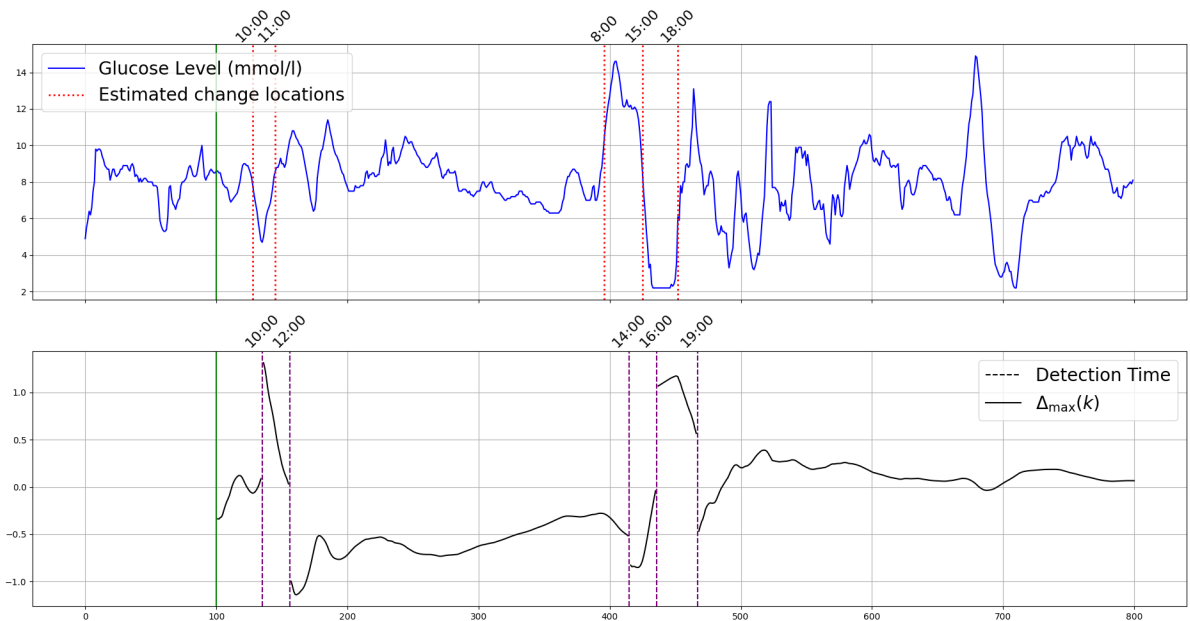}
    \caption{\it Analysis on the blood glucose dataset obtained from Type 1 diabetic patient (as described in Section \ref{sec_11}). The upper graph contains the glucose levels over a two-day period,  where the training sample contains 100 data points, which are observed over the course of 8.33 hours overnight. The green line marks the beginning of the monitoring period. The lower graph shows the progression of $\widehat{\Delta}_{\max} (k)$ multiplied by the sign of the detected deviation from the initial mean. The vertical lines in this plot denote the times when the changes are located.}
    \label{fig:glucose_appendix}
\end{figure}

We continue our discussion of the diabetes dataset previously introduced in Section \ref{sec_11}.  The data was privately acquired from a 29-year-old male with Type 1 Diabetes, who volunteered to share his glucose dataset. Data was collected continuously using body-worn sensors for the purpose of showcasing the use of the methods presented in this paper. The automated closed-loop sensors recorded glucose levels at irregular time intervals, with an average sampling frequency of approximately one reading every five minutes.  The dataset spans from May 2, 2025, at 01:02 CET to May 5, 2025, at 03:26 CET.  The start of the dataset is so chosen in order to ensure that there is no meal time and no physical exertion for the next 8.3 hours thus ensuring that the blood glucose levels remain fairly stable and hence suitable as a baseline period. 

The focus of our analysis is detecting dangerous spikes in glucose concentration. Glucose levels over time are displayed in the upper panel of Figure \ref{fig:glucose_appendix} (this is an extension of Figure \ref{fig:glucose_levelMotivation}). 
In Section \ref{sec_11}, we compared glucose levels to normative lower and upper bounds, that we will discuss in more detail now. 
Values below 4 mmol/L (70 mg/dL) are considered concerningly low (\textit{hypoglycemia}), requiring timely action (\cite{Cryer2009}). Hypoglycemia is associated with symptoms such as shakiness, rapid heartbeat, confusion, dizziness, blurred vision, seizures, and in severe cases, loss of consciousness; severe hypoglycemia is a medical emergency.
Excessively high glucose levels, or \textit{hyperglycemia}, are often defined as values exceeding 10 mmol/L (180 mg/dL) postprandially (after a meal). Due to considerable individual variability in symptom onset, defining severe hyperglycemia is non-trivial; values above 13.3 mmol/L are sometimes used in the literature as this is the threshold where ketoacidosis can set in \cite{Abbas2009}, but much higher values can be reached. The most important symptoms of acute hyperglycemia include excessive thirst, frequent urination, fatigue, blurred vision, and confusion. 
In Figure \ref{fig:glucose_levelMotivation}, we have chosen to display glucose values between 4 and 12 mmol/L as normative. It is important to point out that the patient in this study uses an insulin pump that automatically adapts insulin to glucose levels, preventing any dangerous concentrations for extended periods of time.
 \medskip
 
\textbf{Statistical results} As in the case of the polling data, we apply sequential statistical methods to the glucose dataset, to mimic a real-time monitoring. In Section \ref{sec_11}, we have applied a range of existing monitoring methods (i.e. sequential change point tests) to the glucose dataset and found that all raise alarms very early. A depiction of this is given in Figure \ref{fig:glucose_levelMotivation} and the precise times of rejection are collected in Table \ref{compareBG_otherMeth}. All methods raise an alarm and stop after at most $6\%$ of the monitoring period and all alarms are raised while glucose values are well within the normative range. 

\begin{table}[]
    \centering
    \begin{tabular}{|c|c|}
    \hline
         Method& Change location  \\
         \hline \hline
         HKHS& 11:36:02 ($k=39$)\\
        KD & 11:06:02 ($k=33$)\\
        DG & 11:31:02 ($k=38$)\\
        F & 11:21:02 ($k=36$)\\
        \hline
    \end{tabular}
    \caption{
    Detected change points for the diabetes dataset recorded on 02-05-2025
    by the four  comparative methods given in \cite{horvath:kokoszka:huskova:steinebach:2003} (HKHS), \cite{fremdt:2015} (F), \cite{dette:gossmann:2019} (DG), and \cite{kutta:doernemann:2025} (KD). 
  In the parentheses, we denote the number of observations (\# Obs.) after the start of monitoring at which the change was detected.}
    \label{compareBG_otherMeth}
\end{table}

Next, we apply our new statistical method to the data. As in our analysis of the polling data, we set the nominal level to $\alpha = 5\%$ and $\beta=0.45$ (this is the weight parameter in Algorithm \ref{alg:change_point_estimation} for multiple change point detection). Again, we report the quantity $\widehat{\Delta}_{\max}(k)$ defined in 
\eqref{delta_maxK} and  recall that this value demarcates the boundary where  $H_0(\Delta,k)$ is rejected for any $\Delta<\widehat{\Delta}_{\max}(k)$. Thus, $\widehat{\Delta}_{\max}(k)$ serves as a measure of relevant deviation at any time $k$. In the lower panel of Figure \ref{fig:glucose_appendix}, we display $\widehat{\Delta}_{\max}(k)$ multiplied by the sign of the deviation. Purple vertical lines demarcate times when a change has been detected by  Algorithm \ref{alg:change_point_estimation} and in the upper panel the red vertical lines correspond to the estimated location by Algorithm \ref{alg:change_point_estimation}. We observe that $\widehat{\Delta}_{\max}(k) \neq 0$ almost everywhere, i.e. there exists at most points at least some statistically significant deviation. However, this does not automatically indicate practical relevancy. Indeed, the observed deviations are always less than $1.8$ in absolute terms, indicating a fair level of stability for glucose levels in the patient. Values of $4$ or more might be a cause for concern because this might indicate either hypo- or hyperglycemia. Accordingly, run with any medically reasonable choice of $\Delta$ (say $4$), our method finds no relevant deviation across the entire period and does not trigger a (medically unwarranted) alarm. 
\bigskip

\textbf{Acknowledgments:} We thank the Regional Computing Center of the University of Cologne (RRZK) for providing computing time on their High Performance Computing (HPC) systems: CHEOPS and RAMSES. This research was partially funded by the
Deutsche Forschungsgemeinschaft (DFG);  TRR 391 Spatio-temporal Statistics for the Transition of Energy and Transport (520388526). Tim Kutta's work has been partially funded by AUFF grants 47331 and 47222. Rupsa Basu's work was funded by Deutsche Forschungsgemeinschaft (DFG) by the  project titled: `` \textit{Modeling functional time series with dynamic factor structures and points of impact}",  with project number 511905296. The software implementation and a Python package is found in \href{https://github.com/rupsabasu2020/sequential_relevant_cp}{the Github Repository: (\texttt{sequential\_relevant\_cp})}. 

\small\renewcommand{\baselinestretch}{0.95}
\putbib
\end{bibunit}

\newpage

\setcounter{page}{1}
\section{Appendix}

\begin{bibunit}

\textbf{Some notations}
For two sequences of positive numbers $(a_N)_N, (b_N)_N$, we say that $a_N \lesssim b_N$ if there exists come $C>0$ that does not depend on $N$ and for which 
\begin{align}
    a_N \leq C b_N\qquad \forall N \in \mathbb{N}.
\end{align}
An event $\mathcal{A}$ on the probability space is said to hold with high probability whenever $\mathbb{P}(\mathcal{A})=1-o(1)$.

\subsection{Proof of Theorem \ref{thm1}}
\begin{proof} To start the proof, recall the definition of the detector $\widehat{\Gamma}(k) $
in \eqref{e:gam} and note that in the following we will implicitly assume that $k>N$ whenever $k$ appears. The detector involves the function $\widehat{\Psi}$, which is defined in \eqref{e:def:psi}. Now, using the triangle inequality and the strong approximation in \eqref{Ass1}, part i), we obtain

\begin{align} \label{e:Gam1}
    \widehat{\Gamma}(k) = &\frac{\sqrt{N}}{k} \bigg( \Big|\frac{k-\hat k}{N}
W(N)- [W(k)-W(\hat k)]+f(\hat k, k)\Big|\\
&\qquad\quad-(k-\hat k)\Delta\bigg)+o_P(1):=\widehat{\Gamma}_1(k)+o_P(1), \nonumber
\end{align}
where
\begin{align} \label{e:deff}
f(j, k) = \sum_{n=j+1}^k(\mu^{(1)}-\mu_n) . 
\end{align}
The object  $\widehat{\Gamma}_1$ is defined in the obvious way.

\textbf{Step 1: Splitting up the sup}\\
We will now further analyze the object $\widehat{\Gamma}_1$ defined in \eqref{e:Gam1}. More precisely, with $T_N$ being the growing threshold from Proposition \ref{lem:cp}, we note that
\begin{align}
\label{pb1}
    \sup_k\widehat{\Gamma}_1(k)=\max\{\sup_{k \leq NT_N}\widehat{\Gamma}_1(k),\sup_{k>NT_N}\widehat{\Gamma}_1(k)\}.
\end{align}
The main challenge is the asymptotic analysis of  $\sup_{k \leq NT_N}\widehat{\Gamma}_1(k)$ - the following Steps 2-4. The analysis of $\sup_{k>NT_N}\widehat{\Gamma}_2(k)$ is much simpler. Indeed, from the law of iterated logarithm it follows that 
\[
    \sup_{k > NT_N}\frac{\sqrt{N}}{k}|W(k)-W(\hat k)|=o_P(1),
\]
while the deterministic terms $f(\hat k,k)-(k-\hat k)\Delta$ are upper bounded by 0. Consequently
\begin{align}    
\label{e:Tailbound}
\sup_{k>NT_N}\widehat{\Gamma}_1(k) \leq \frac{\sqrt{N}}{k}\frac{k-\hat k}{N}|W(N)|+o_P(1)\leq N^{-1/2}|W(N)|+o_P(1)
\end{align}
\textbf{Step 2: Removing $k$ for which change point localization is inconsistent}
We define blocks of time points $k$ 
\begin{align}
S_1&:=\{N,N+1,\cdots,\lfloor N\theta_1\rfloor+\lfloor N^{\zeta} \rfloor-1\}\\
S_i&:=\{\lfloor N\theta^{(i)}\rfloor,\lfloor N\theta^{(i)}\rfloor+1,\cdots,\lfloor N\theta^{(i)}\rfloor+\lfloor N^{\zeta} \rfloor-1\}\\
L_i&:=\{\lfloor N\theta^{(i)}\rfloor+\lfloor N^{\zeta}\rfloor,\cdots, \lfloor N\theta^{(i+1)}\rfloor-1\}.
\end{align}
The above blocks form a disjoint decomposition of the numbers $\{N, N+1,\cdots \}$. The blocks $S_i$ are small blocks right after a change (when the change is theoretically hard to detect) and the large blocks $L_i$ denote the time span that is sufficiently separated from the most recent change $\theta^{(i)}$ to ensure consistent detection. 
Now, we want to show that 
\begin{align}
\label{pb2}
    \sup_{k \leq NT_N}\widehat{\Gamma}_1(k)=\sup_{i: \theta^{(i)}<T_N}\sup_{k \in L_i}\widehat{\Gamma}_1(k)+o_P(1)~.
\end{align}
To that end we note that for all $k \in S_i$ for which $k \leq NT_N$ the estimator  $\hat k$ behaves (by Proposition \ref{lem:cp}) with high probability according to one of the following two scenarios:
\begin{enumerate}
    \item[1)]$|\hat k - \lfloor N\theta_{i-1}\rfloor |\leq N^{\zeta-1}$ in the case that $|\hat \Theta(k)|<|\Theta(k)|$.
    \item[2)]$|\hat k-\lfloor N\theta^{(i)} \rfloor||\leq N^{\zeta-1} $ in the case that $|\hat \Theta(k)|=|\Theta(k)|$.
\end{enumerate}
Using  Lemma \ref{lem:BrownianCont}, one has that
\begin{align}
   \max_{i: \theta^{(i)} <T_N}\sup_{k \in S_i} \min\Big\{|\widehat{\Gamma}_1(k)-\widehat{\Gamma}_1(\lfloor N\theta^{(i)} \rfloor-1)|,|\widehat{\Gamma}_1(k)-\widehat{\Gamma}_1(\lfloor N\theta^{(i)} \rfloor+N^{\zeta})|\Big\}=o_P(1)~.
\end{align}
which yields equation \eqref{pb2} because $\lfloor N\theta^{(i)} \rfloor-1$ and $\lfloor N\theta^{(i)} \rfloor+N^{\zeta}$ are contained in $L_{i-1}$ and $L_i$, respectively. \\
\textbf{Step 3: Replacing $\hat k$ by $\check k$}\\
Recall the defintion of $\check k$ in \eqref{e:defk*}. We now want to replace $\widehat{\Gamma}_1(k)$ by
\begin{align}
    \tilde \Gamma_1(k):=\frac{\sqrt{N}}{k} \bigg( \Big|\frac{k-\check k}{N}
W(N)- [W(k)-W(\check k)]+f(\check k, k)\Big|-(k-\check k)\Delta\bigg)
\end{align}
in the right hand side of equation \eqref{pb2}. More precisely, we aim to show
\begin{align}
    \label{pb9}
     \sup_{k \leq NT_N}\widehat{\Gamma}_1(k)=\sup_{i, \theta^{(i)}<T_N}\sup_{k \in L_i}\tilde \Gamma_1(k)+o_P(1).
\end{align}
To that end note that with high probability it holds that, simultaneously for all those $k$ with $k \leq NT_N$ and $k \in L_i$ for some $i$ we have 
\begin{align}
    |\hat k -\check k|\leq N^{\zeta-1}~.
\end{align}
This claim follows by Proposition \ref{lem:cp}.
Consequently we obtain by Lemma \ref{lem:BrownianCont} that
\begin{align}
    \sup_{i: \theta^{(i)}<T_N}\sup_{k \in L_i}\frac{\sqrt{N}}{k}|W(\hat k)-W(\check k)| \leq O_P(1) \sqrt{N^{\zeta-1}}(1+\log(N))~.
\end{align}
For the application of the lemma set $t_0 = NT_N$ and $C_1=1$, and also notice that $\log(N) \sim \log(NT_N)$ because $T_N$ is polynomially growing in $N$.
Similarly we have
\[
    \sup_{i: \theta^{(i)}<T_N}\sup_{k \in L_i}\frac{\sqrt{N}}{k}|\check k-\hat k|\Delta\leq O_P(1)\frac{N^{\zeta-1/2}}{k}~.
\]
Analogous arguments for the other terms in $\widehat{\Gamma}_1(k)$ and $\tilde \Gamma_1(k)$ yield the desired statement.\\
\textbf{Step 4: Discarding deviations smaller than $\Delta$}\\
Recall the definitions of the function $f$ in \eqref{e:deff}, the function $\Psi$ in \eqref{e:defpsinew} and the parameter $\check{\theta}$  in \eqref{e:defk*}. We note that
\begin{align}
\label{pb4}
    |f(\check k,k)|-(k-\check k)\Delta\leq -(k-\check k)\Big(|\Psi(1)-\Psi(\check \theta)|-\Delta\Big)~.
\end{align} 
Further, we observe by Lemma \ref{lem:BrownianCont} that
\begin{align}
\label{pb5}
    \frac{k-\check k}{N}|W(N)|&=\frac{k-\check k}{\sqrt{N}}O_P(1),\\
    \label{pb6}
    |W(k)-W(\check k)|&\leq \sqrt{(k-\check k)\log(N)}O_P(1),
\end{align}
uniformly in $k \in \bigcup_{\theta^{(i)} \leq T_N} L_i$.
Combining \eqref{pb4}-\eqref{pb6}  yields that 
\begin{align*}
    &\sup_{\delta_i<\Delta-\epsilon, \theta^{(i)}<T_N, }\sup_{k \in L_i}\tilde \Gamma_1(k)\\
    \leq &  \sup_{\delta_i<\Delta-\epsilon, \theta^{(i)}<T_N, }\sup_{k \in L_i}
   \frac{\sqrt{N}}{k} \bigg( \Big|\frac{k-\check k}{N}
W(N)\Big|- \Big|W(k)-W(\check k)\Big|+\Big|f(\check k, k)\Big|-(k-\check k)\Delta\bigg)\\
\leq &\sup_{ \substack{k\leq NT_N\\ k-\check k \geq N^\zeta}}\Bigg(\frac{ O_P(1)\sqrt{N}}{k} \Big(\frac{k-\check k}{\sqrt{N}}+\sqrt{(k-\check k)\log(N)}\Big) -(k-\check k)\epsilon\Bigg)
= o_P(1).
\end{align*}
In the last inequality, we have used that $k \in L_i$ (for some $i$), which implies that $k-\check k \geq N^{\zeta}$.\\
Now, combining the identities \eqref{pb2} and \eqref{pb9}, this already yields the fact that $\sup_{k\leq NT_N}\widehat{\Gamma}(k)= o_P(1)$ if $\delta<\Delta$ - combining this with \eqref{e:Tailbound} yields the upper part of eq. \eqref{pb12} in Theorem \ref{thm1}.

We are left with the case that $\delta=\Delta$. Define
\begin{align} \label{e:defAN}
     \mathfrak{A}_N:= \Big\{j: |\delta_j|>\Delta - \frac{\log(N)}{\sqrt{N}}\Big\}.
\end{align}  
Note that for those $k \in L_i$  for which $i \in \mathfrak{A}_n$ we observe that due to \eqref{pb5} and \eqref{pb6} we have, uniformly over $k \in \cup_{i, \theta^{(i)}<T_N} L_i$, that
\begin{align}
\label{pb3}
    \text{sign}\Big(\frac{k-\check k}{N}W(N)- [W(k)-W(\check k)]+f(\check k, k)\Big)=\text{sign}(f(\check k,k))=:\check s.
\end{align}
Here the first equality holds with high probability.
Using this we now obtain that the following chain of inequalities holds with high probability, arguments for each line are provided below. 
\begin{align}    
     &\sup_{i \in \mathfrak{A}_N , \theta^{(i)}<T_N, }\sup_{k \in L_i}\tilde \Gamma_1(k)\\
     \label{pb20}
     =& \sup_{i \in \mathfrak{A}_N , \theta^{(i)}<T_N, }\sup_{k \in L_i}  \frac{\sqrt{N}}{k} \bigg(\check s\Big(\frac{k-\check k}{N} W(N)- [W(k)-W(\check k)]+f(\check k, k)\Big)\\
     &  \qquad \qquad \qquad \qquad \qquad \qquad-(k-\check k)\Delta\bigg)\\
     \label{pb21}
     = &\sup_{i \in \mathfrak{A}_N, \theta^{(i)}<T_N}\sup_{ N\theta^{(i)}  \leq k \leq  N\theta^{(i+1)} }\frac{\sqrt{N}}{k} \bigg( s_i\Big(\frac{k- N\theta^{(i)}}{N}W(N)- [W(k)-W(N\theta^{(i)})]\\
      & \qquad \qquad \qquad \qquad \qquad \qquad +f(\check k, k)\Big)-(k-\check k)\Delta\bigg)+o_P(1)\\
     \label{pb22}
 = &\sup_{i \in \mathfrak{A}, \theta^{(i)}<T_N, }\sup_{ N\theta^{(i)}  \leq k \leq  N\theta^{(i+1)} } \frac{\sqrt{N}}{k}   s_i\Big(\frac{k-\check k}{N}
W(N)- [W(k)-W(N\theta^{(i)})]\Big)+o_p(1)\\
\label{pb23}
= &\sup_{i \in \mathfrak{A}}\sup_{N\theta^{(i)}  \leq x \leq N\theta^{(i+1)} }\frac{\sqrt{N}}{x} s_i\Big(\frac{x-N\theta^{(i)}}{N}W(N)- [W(x)-W(N\theta^{(i)})]\Big)+o_P(1)\\
\label{pb8}
=&L+o_P(1).
\end{align}

\begin{enumerate}
    \item \textbf{Equality   \eqref{pb20} $=$ \eqref{pb21}} \\
    The index set of the inner supremum changes (it gets bigger). This follows by the same continuity argument that was used in Step 2
    \item \textbf{Equality \eqref{pb21} $=$ \eqref{pb22}}. \\
    Define
            \begin{align}
                        i_{\min} =\min \Big(\mathfrak{A}_N\setminus \{j \ |\ \delta_i=\Delta\}\Big)=\min\Big(\mathfrak{A}_N \setminus \mathfrak{A}\Big)~.
            \end{align}
    There are two possibilities: Either eventually $i_{\min}=\infty$ in the case that there are only finitely many elements  in $\mathfrak{A}_N$ for which $\delta_i<\Delta$, in that case we are done. Otherwise we have that $i_{\min} \rightarrow \infty$ in the case that there are infinitely many such elements. In the latter case we will show that
    \begin{align}
    \label{pb14}
        \sup_{i \geq i_{\min}}  \sup_{N\theta^{(i)}  \leq x \leq N\theta^{(i+1)} }&\frac{\sqrt{N}}{x} s_i\Big(\frac{x-N\theta^{(i)}}{N}W(N)- [W(x)-W(N\theta^{(i)})]\Big)\\
         & +s_i f(N\theta^{(i)}, x)-(x-N\theta^{(i)})\Delta\bigg)=o_p(1)~,        
    \end{align}
    which implies \eqref{pb22}. 
    For the proof of equation \eqref{pb14} we observe that by Assumption \ref{Ass1}, part iii) we have
    \begin{align}
    \label{pb17}
         \sup_{i \geq i_{\min}}\sup_{N\theta^{(i)}  \leq x \leq N\theta^{(i+1)} }\frac{(x-N\theta^{(i)})}{x}\frac{|W(N)|}{\sqrt{N}}&=\sup_{i \geq i_{\min}}\frac{(\theta^{(i+1)}-\theta^{(i)})}{\theta^{(i+1)}}\frac{|W(N)|}{\sqrt{N}}\\
         &=o_P(1)~.
    \end{align}
    For the increment terms in \eqref{pb14} we observe that
    \begin{align}
         &\sup_{i \geq i_{\min}} \sup_{N\theta^{(i)}  \leq x \leq N\theta^{(i+1)} }\frac{\sqrt{N}}{x}\Big|W(x)-W(N\theta^{(i)})\Big|\\
        \overset{d}{=}& \sup_{i \geq i_{\min}}\sup_{\theta^{(i)} \leq x \leq \theta^{(i+1)}}\frac{\sigma|W_i(x-\theta^{(i)})|}{x} \\
        \le& \sup_{i \geq i_{\min}}\sup_{\theta^{(i)} \leq x \leq \theta^{(i+1)}}\frac{\sigma|W_i(x-\theta^{(i)})|}{\theta^{(i)}} \\\overset{d}{=} &\sup_{i \geq i_{\min}}\sup_{ 0\leq x \leq \theta^{(i+1)}-\theta^{(i)} }\frac{\sigma \sqrt{\theta^{(i+1)}-\theta^{(i)}}|W_i(x/(\theta^{(i+1)}-\theta^{(i)}))|}{\theta^{(i)}} \\
        \overset{d}{=}&\sup_{i \geq i_{\min}}\sup_{0 \le x \le 1}\frac{\sigma\sqrt{\theta^{(i+1)}-\theta^{(i)}}|W_i(x)|}{\theta^{(i)}}\\
        \le & \sup_{i \geq i_{\min}}\sup_{0 \le x \le 1}\frac{\sigma\sqrt{\theta^{(i+1)}-\theta^{(i)}}|W_i(x)|}{\sqrt{\theta^{(i)}}\sqrt{\theta^{(i)}}} = o(1) \sup_{i \geq i_{\min}}\sup_{0 \le x \le 1}\frac{|W_i(x)|}{\sqrt{\theta^{(i)}}}\\
        \le  & o(1) \sup_{i \geq i_{\min}}\sup_{0 \le x \le 1}\frac{|W_i(x)|}{\sqrt{i \underline{C}}} \le o(1) \sup_{i \ge i_{\min}}\frac{\sup_{0 \le x \le 1}|W_i(x)|}{\sqrt{i \underline{C}}} \\
        \le &o(1)\max_{1\le i <\infty}\frac{\sup_{0 \le x \le 1}|W_i(x)|}{\sqrt{i \underline{C}}}.
    \end{align}
    Using the union bound and that $\sup_{0 \le x \le 1}|W_i(x)|$ is subgaussian yields that 
    \begin{align}
         \max_{1\le i <\infty}\frac{\sup_{0 \le x \le 1}|W_i(x)|}{\sqrt{i}}
    \end{align}
    is a.s. bounded (using a Borel-Cantelli argument). We hence conclude that
    \begin{align}
    \label{pb16}
        \sup_{i \geq i_{\min}} \sup_{N\theta^{(i)}  \leq x \leq N\theta^{(i+1)} }\frac{\sqrt{N}}{x}\Big|W(x)-W(N\theta^{(i)})\Big|=o_p(1).
    \end{align}
    Lastly we also have on account of the fact that $\delta\leq \Delta$ that
    \begin{align}
    \label{pb18}
        \sup_{i \geq i_{\min}} \sup_{N\theta^{(i)}  \leq x \leq N\theta^{(i+1)} }\frac{\sqrt{N}}{x}(s_if(N\theta^{(i)},x)-(x-N\theta^{(i)})\Delta)=0  ~.   
    \end{align}    
    Combining  \eqref{pb17}, \eqref{pb16} and \eqref{pb18} yields \eqref{pb14}.
    
    \item \textbf{Equality  \eqref{pb22}$=$\eqref{pb23}}\\
    One first uses a continuity argument as in Step 2 to transition from $k \in \mathbb{N}$ to $x \in \R$. The statement then follows because $T_N\to \infty$.
    \item \textbf{Equation \eqref{pb23}$=$\eqref{pb8}}\\
    This follows by using the basic properties of Brownian motion.
\end{enumerate}



\end{proof}

\begin{lemma}[\cite{chevallier2023}]
\label{lem:BrownianCont}
The random variable
    \begin{align}
\label{pb10}
    M_B=\sup_{0<s<t<\infty}\frac{|W(t)-W(s)|}{\sqrt{(t-s)(1+\log(\frac{t}{t-s})+\epsilon|\log(t)|)}}
\end{align}
has square exponential moments for any $\epsilon>0$. In particular we have for any   $t_0 >0$ and uniformly over all $x,y\in [0,t_0]$ with $|x-y|\geq C_1$ that
\begin{align}
    |W(x)-W(y)|\leq C_2M_B\sqrt{|x-y|\log(t_0)}~.
\end{align}
Here the constant $C_2$ only depends on $C_1$. 
\end{lemma}

\subsection{Proof of Proposition  \ref{l2}}

\begin{proof}

We proceed in two steps. We first show that
\[
    \max\{B, \hat L_N\}=\max\{B,L\}+o_P(1),
\]
and then establish
\[
    \max\{B,\hat L_N\} \leq \sup_{k>N}\hat L_N(k)+o_p(1)~.
\]
Combining the two yield the desired result. \\

    \textbf{Showing that $\max\{B, \hat L_N\}=\max\{B,L\}+o_P(1)$}\\
    Note that $(\mathfrak{A}\cap\{i| \theta^{(i)}\leq T_N\}) \subset \hat{\mathfrak{A}}_N(NT)$ holds with high probability. This follows by noting that combining Assumption \ref{Ass1} i) and Proposition \ref{lem:cp} (choose $\zeta$ and $\kappa$ so that $\zeta+\kappa<1/2$) implies that
    \begin{align}
        \sup_{\theta^{(i)} \leq T_N}\sqrt{N(\theta^{(i+1)}-\theta^{(i)})}\Big|(\hat \mu^{(i)}-\mu^{(i)})-\frac{W(N\theta^{(i+1)})-W(N\theta^{(i)})}{N(\theta^{(i+1)}-\theta^{(i)})}\Big|=o_P(1).
    \end{align}
    In particular one has by Lemma \ref{lem:BrownianCont} that 
    \begin{align}
        \sup_{\theta^{(i)} \leq T_N}|\hat \mu^{(i)}-\mu^{(i)}| \leq \frac{\sqrt{\log(N)}}{\sqrt{N\underline{C}}}O_P(1),
    \end{align}
    which implies that 
    \begin{align}
          \sup_{\theta^{(i)} \leq T_N}\Big[|\hat \mu^{(1)}-\hat \mu^{(i)}| -\delta_i\Big] \geq - \frac{\sqrt{\log(N)}}{\sqrt{N}}O_P(1),
    \end{align}
    and therefore yields $(\mathfrak{A}\cap\{i| \theta^{(i)}\leq T_N\}) \subset \hat{\mathfrak{A}}_N(NT_N)$ with high probability, as claimed (definition of $\hat{\mathfrak{A}}_N$ is found in \eqref{e:def:AN}). \\
    
    Recalling equations \eqref{pb22}-\ref{pb8} we have
    \[
    L=\sup_{i \in \mathfrak{A}, \theta^{(i)}<T_N, }\sup_{ N\theta^{(i)}  \leq k \leq  N\theta^{(i+1)} } \frac{\sqrt{N}}{k}   s_i\Big(\frac{k-\check k}{N}
W(N)- [W(k)-W(N\theta^{(i)})]\Big)+o_p(1)~.
    \]
    We now replace $\mathfrak{A}$ by the superset $\hat{\mathfrak{A}}_N(NT_N)$ and also replace the population versions of changes and signs by their empirical counterparts to obtain that
    \begin{align}
        \hat L_N\geq L+o_P(1).
    \end{align}
    For the reverse inequality we simply mimic the proof of equation \eqref{pb22} to discard all changes not equal to $\Delta$. The same continuity argument as for the other direction (to swap empirical change points and signs for their population versions) then yields $\hat L_N=L+o_P(1)$ and consequently also $\max\{B, \hat L_N\}=\max\{B,L\}+o_P(1)$.\\
    
    \textbf{Showing that $\max\{B,\hat L_N\} \leq \hat L_N(k)$}\\
    Note that $B=W(N)/\sqrt{N}$ by the proof of Theorem \ref{thm1}.
    Define the running index 
    \begin{align}
        \hat i(k)=\begin{cases}
            \max{\hat{\mathfrak{A}}_N(k)} \quad &\hat{\mathfrak{A}}_N(k) \neq \emptyset \\
            1  & \hat{\mathfrak{A}}_N(k) =\emptyset 
        \end{cases}
    \end{align} 
    If the first sup in the definition of $\hat L_N$ is taken on for some $i \leq \hat i(k)$ and $\Big| \hat \Theta(k)\Big|\geq 2$ we have $\hat L_N=\hat L_{N,1}(k)$ (recall the definitions of these objects in eq. \eqref{e:def:Lh} and above). If $| \hat \Theta(k)|\leq 1$ we instead have $\hat L_N \leq \hat L_{N,2}(k)$  In the case that the supremum is not attained for $i \leq \hat i(k)$ we know that it has to be attained for some $i \geq \max (\hat \Theta(k))$. One can verify by elementary calculations that for each $i\geq \max(\hat \Theta(k))$ with $i \in \hat{\mathfrak{A}}_N(NT_N)$ and any $x\geq \hat \theta^{(i)}$ we have
    \begin{align}
        s_i\frac{\sqrt{N}}{Nx}\Big((x-\hat \theta^{(i)})W(N)-\Big[W(Nx)-W(N\hat \theta^{(i)})\Big]\Big) \leq \hat L_{N,2}(k)
    \end{align}
    which yields
    \begin{align}
        \hat L_N\leq \hat L_{N,2}(k)
    \end{align}
    in this case. Either way $\hat L_N$ is upper bounded by $\hat L_N(k)$. 
    Finally, we show that almost surely
    \begin{align}
        B \leq \hat L_{N,2}(k).
    \end{align}
for this purpose, recall the definition of $\hat L_{N,2}(k)$ in \eqref{hd1}. We make $\hat L_{N,2}(k)$ smaller by choosing a specific pair $(x, \theta)$ (rather than maximizing over such pairs) and (for a fixed outcome) obtain for any sufficiently large $\theta$ 
\begin{align*}
    \hat L_{N,2}(k) \ge \frac{\sqrt{N}}{N\theta^2}\Bigg|(\theta^2-\theta)W(N)-\Big[W(N\theta^2)-W(N\theta)\Big]\Bigg|=:\tilde L(\theta)
\end{align*}
by setting $x=\theta$. Now, using the law of the iterated logarithm and letting $\theta \to \infty$ we obtain a.s.
\[
\tilde L(\theta) \to \frac{|W(N)|}{\sqrt{N}}=|B|.
\]
    This then yields the desired result.
\end{proof}

\subsection{Proof of Proposition \ref{lem:cp}}

\textbf{Definitions} For a stochastic process $P: [0, T_N] \to \mathbb{R}$ we first introduce the second-order difference functional 
\[
\mathbf{R}[P](k+N,\ell):=\sqrt{N}\{P([N+k]/N)-2P([N+k-\ell]/N)+P([N+k-2\ell]/N)\}.
\]
Correspondingly, we define the mean difference functional
\[
\mathbf{M}(k+N,\ell):= \sum_{n=N+k-2\ell+1}^{N+k-\ell}\mu_n - \sum_{n=N+k-\ell+1}^{N+k}\mu_n.
\]
Next, we introduce the partial sum process of model errors
\begin{align*} 
P_{N}(x):= N^{-1/2} \sum_{i=1}^{\lfloor x N \rfloor} \varepsilon_{i}+N^{-1/2} \big[x-(\lfloor N x \rfloor/N )\big] \varepsilon_{\lceil N x \rceil}, \qquad x \in [0,T_N].
\end{align*}
Furthermore, we define the set of all permissbile pairs $(\ell, k)$ as
\[
\mathcal{S}:= \{(\ell,k): k, \ell \ge 1,\,\, N+k-2\ell\ge 0\}.
\]

\textbf{Construction of the event $\mathcal{E}_N$}
With these definitions in place, we can rewrite the test statistic as follows
\begin{align*}
    \gamma(\ell,k+N)=  w_\beta(\ell,k+N) \cdot \Big| \mathbf{R}[P_N](k+N,\ell)+\mathbf{M}(k+N,\ell)\Big|.
\end{align*}
In order to construct the event $\mathcal{E}_N$, we focus on the stochastic part of the above (without $\mathbf{M}(k+N,\ell)$) and define it as 
\begin{align*}
    \gamma'(\ell,k+N):=  w_\beta(\ell,k+N) \cdot \Big| \mathbf{R}[P_N](k+N,\ell)\Big|.
\end{align*}
Using the same arguments as in the proof of Lemma 3.3 of \cite{kutta:doernemann:2025}, we can show that
\begin{align}
&\sup_{(\ell, k) \in \mathcal{S}}| \gamma'(\ell,k+N)| \overset{d}{\to} V,\qquad \textnormal{with the weak limit defined as }\nonumber\\
V:=&\sup_{\substack{0 \le x < y < \infty \\ 0 \le 1 + y - 2x }}
\frac{|W(1+y)-2W(1+y-x)+W(1+y-2x)|}{(1+y)^{1-\beta}x^\beta\log(2+y)}. \label{e:defV}
\end{align}
We point out that for this result to hold, Assumption \ref{Ass1} part ii) is needed (Assumption 3.2, part ii) of \cite{kutta:doernemann:2025}).
Now, we define the set
\[
\mathcal{E}_N:=\Big\{\sup_{(\ell, k) \in \mathcal{S}} \gamma'(\ell,k+N) \le C_{CP} \log(N)\Big\}
\]
and it follows by the weak convergence in \eqref{e:defV} that 
\[
\mathbb{P}(\mathcal{E}_N) \to 1.
\]

\textbf{Proof of the lemma}  Let $C_T>0$ be a constant such that $T_N \le C_T N^{\kappa}$ (this is possible by construction of $T_N$). Also notice, that by the assumptions of the lemma on the constant $\zeta>0$, it holds that 
\[
c_\zeta := 1/2+\zeta - (1+\kappa)(1-\beta)+\zeta \beta
\]
is positive. Recall the definition of the constant $\underline{C}>0$ from Assumption \ref{Ass1}, part iii) and the constant $C_{CP}$ from the definition of the set $\widehat{\Theta}(k)$ in \eqref{e:defhT}. Then, 
we assume that the number $N_0$ in the formulation of this lemma is chosen large enough such that simultaneously the following inequalities hold for all $N\ge N_0$:
\begin{itemize}
    \item[(a)] $\lfloor N^\zeta \rfloor < \min_i |\lfloor \theta^{(i)} N \rfloor - \lfloor \theta^{(i+1)} N \rfloor|/4$, $\quad\lfloor N^\zeta \rfloor<N/4$.
    \item[(b)] $1<T_N,$ $\quad$ $3C_T N^\kappa \le N$.
    \item[(c)] 
    $
    [\underline{C}/(4C_T)] N^{c_\zeta}>2 C_{CP} \log^2(N).
    $
\end{itemize}
Now, the remainder of the proof follows by successively investigating all changes $\theta_1<\cdots <\theta_{K_N}<T_N$. The case where $K_N=0$ is simpler than the rest and hence neglected. We focus exemplarily on the first change only, but all bounds that we derive hold uniformly for any of the changes and hence using the same arguments iteratively for the successive changes is trivial. We need to consider two cases: First, the period before the change has happened, and second, the period shortly after the change. Beginning with the episode before the change,
we start by showing that, conditionally on $\mathcal{E}_N$
 \begin{align} \label{e:T=0}
      0= |\widehat{\Theta}(k)| = |\Theta(k)|
 \end{align}
for $k=N+1,\cdots ,\lfloor N \theta_1 \rfloor -1$. Since, $\widehat{\Theta}(k)$ is monotonically growing in $k$ we can restrict ourselves to the case $k'= \lfloor N \theta_1 \rfloor -1$. The algorithm will only add elements to $\widehat{\Theta}(k')$ if the condition in line 9 is fulfilled, which in turn is implied by
\[
\max_{(\ell, k) \in \mathcal{S}, k \le k'} \gamma(\ell,k+N)>C_{CP}\log(N).
\]
However, since no change has occurred for $k<k'$, we have 
\[
\max_{(\ell, k) \in \mathcal{S}, k \le k'} \gamma(\ell,k+N) 
= \max_{(\ell, k) \in \mathcal{S}, k \le k'}  \gamma'(\ell,k+N) \le C_{CP}\log(N),
\]
where the last inequality holds on the event $\mathcal{E}_N$. This shows \eqref{e:T=0} for $k=N+1,\cdots ,\lfloor N \theta_1 \rfloor -1$. Next, we consider values of $k$ in $\lfloor N \theta_1 \rfloor, \lfloor N \theta_1 \rfloor+1,\cdots ,\lfloor N \theta_1 \rfloor+h_N$ and since the set $\widehat{\Theta}(k)$ is monotonically growing in $k$, we can focus on the endpoint $\lfloor N \theta_1 \rfloor+h_N$. Here $h_N := \lfloor N^\zeta \rfloor$ and, since no change has been detected so far, in the algorithm we still have $\bar k=N$. We consider the condition in line 9, and specifically see that
\[
\max_{\ell  \in \mathcal{S}(k,\bar k)} \gamma(\ell, \lfloor N \theta_1 \rfloor+h_N) \ge \gamma(h_N, \lfloor N \theta_1 \rfloor+h_N).
\]
We analyze $\gamma(h_N, \lfloor N \theta_1 \rfloor+h_N)$ and observe the bound
\begin{align*}
& \gamma(h_N,\lfloor N \theta_1 \rfloor+h_N)\\
\ge & w_\beta(h_N,\lfloor N \theta_1 \rfloor+h_N) \cdot |\mathbf{M}(\lfloor N \theta_1 \rfloor+h_N,h_N)| - \sup_{(\ell, k) \in \mathcal{S}} w_\beta(\ell,k+N) \cdot \Big| \mathbf{R}[P_N](k+N,\ell)\Big| \\
=& w_\beta(h_N,\lfloor N \theta_1 \rfloor+h_N) \cdot|\mathbf{M}(\lfloor N \theta_1 \rfloor+h_N,h_N)| - \sup_{(\ell, k) \in \mathcal{S}} \gamma'(\ell,k+N)\\
\ge & w_\beta(h_N,\lfloor N \theta_1 \rfloor+h_N) \cdot|\mathbf{M}(\lfloor N \theta_1 \rfloor+h_N,h_N)| - C_{CP} \log(N).
\end{align*}
Analyzing the first term shows that 
\[
|\mathbf{M}(\lfloor N \theta_1 \rfloor+h_N,h_N)| = h_N |\mu^{(1)}-\mu^{(2)}| \ge \underline{C} h_N,
\]
where we have used the lower bound from Assumption \ref{Ass1}, part iii) in the second step. Thus, plugging in the definition of 
$w_\beta$ in eq. \eqref{e:wbeta}, we get 
\begin{align}
    &w_\beta(h_N,\lfloor N \theta_1 \rfloor+h_N) \cdot|\mathbf{M}(\lfloor N \theta_1 \rfloor+h_N,h_N)|\nonumber \\
    \ge & \frac{\underline{C} N^{1/2} h_N }{(\lfloor N \theta_1 \rfloor+h_N)^{1-\beta}h_N^{\beta} \log(1+(\lfloor N \theta_1 \rfloor+h_N)/N)}.\label{e:btrans}
\end{align}
We analyze further the right side, beginning with the factors in the denominator. Therefore, notice that $h_N<\lfloor N \theta_1 \rfloor$  (condition (a)), $\theta_1>1$ and hence
\[
(\lfloor N \theta_1 \rfloor+h_N)^{1-\beta} \le 2 (\theta_1 N)^{1-\beta} < 2 (T_N N)^{1-\beta} \le 2 C_T N^{(1+\kappa)(1-\beta)},
\]
where (as we recall) $C_T>0$ is a constant that we have fixed, s.t.  $T_N \le C_T N^{\kappa}$. Next, 
$h_N \le N^\zeta$ by definition which together with (b) yields
\[
1+ \frac{\lfloor N \theta_1 \rfloor+h_N}{N} \le 1+2T_N \le 3T_N \le 3 C_T N^\kappa \le N
\]
and hence
\[
\log(1+(\lfloor N \theta_1 \rfloor+h_N)/N) \le \log(N).
\]
Finally, to analyze the numerator, we simply notice that $h_N \ge N^\zeta/2$. Thus, we can lower bound (and simplify) the right side of \eqref{e:btrans} by 
\[
 B:= \frac{\underline{C}}{4C_T} N^{c_\zeta} \log^{-1}(N), \qquad \textnormal{where} \quad  c_\zeta = 1/2+\zeta - (1+\kappa)(1-\beta)+\zeta \beta.
\]
Now, we want to check whether $B>2 C_{CP} \log(N)$, in which case we would have
\[
\gamma(h_N,\lfloor N \theta_1 \rfloor+h_N) >C_{CP} \log(N)
\]
by our above considerations. This however is true according to (c).  Circling back to line 9 of the algorithm, this means that for at least one value of $r=\lfloor N \theta_1 \rfloor, \lfloor N \theta_1 \rfloor+1,\cdots ,\lfloor N \theta_1 \rfloor+h_N$, the if-condition in line 9 is met, and a change is detected. We call the smallest value $r'$ and we have $\Theta(r') = \widehat{
\Theta}(r')=1$.

Now, the proof can be completed by iteratively going through all remaining changes. If none exist, it is easy to show that the condition in line 9 is never again satisfied, since on $\mathcal{E}_N$, we have (for any $r'<r\leq k$)
\[
\gamma:= \max_{\ell  \in \mathcal{S}(r,\bar k)} \gamma(\ell, r) \le \sup_{(\ell, k') \in \mathcal{S}} \gamma'(\ell,k'+N) \le C_{CP} \log(N).
\]
If a second change occurs, one can by the same arguments as before show that
\[
      1= |\widehat{\Theta}(k)| = |\Theta(k)|
\]
for $k=r', r'+1,\cdots ,\lfloor N \theta_2 \rfloor-1$. And then, as before, we can show that a change is located at a position $r''$ in  $\lfloor N \theta_2 \rfloor, \lfloor N \theta_2 \rfloor+1,\cdots ,\lfloor N \theta_2 \rfloor+h_N$. The argument is identical to those before, since we have always applied uniform bounds that did not depend on $\theta_1$, but rather can be applied to all change point locations (by using $\theta^{(i)} \le T_N$). Proceeding like this then completes the proof. \hfill $\square$

\subsection{Motivating the definition of $\tilde{\mathfrak{A}}_N(k)$}
\label{sec:motiv1}
In this subsection we motivate the definition of $\tilde{\mathfrak{A}}_N(k)$ when testing for multiple choices of $\Delta$ at once. To that end let us recall its definition given by 
\begin{align}
     \tilde{\mathfrak{A}}_N(k):=& \Big\{i: |\hat \Psi(1)-\hat \Psi(N\hat \theta^{(i+1)})|\\
    &\quad\quad\geq \max_{l:\hat \theta^{(l+2)} \leq k/N} |\hat\Psi(1)-\hat \Psi(N\hat \theta^{(l+1)})|-\frac{\log(N)}{\sqrt{N}}, \hat \theta^{(i+2)}\leq k/N\Big\}
\end{align}
and compare it to that of $\hat{\mathfrak{A}}_N(k)$ for different choices of $k$ and $\delta$. For the sake of readability we will neglect mentioning "with high probability" for statements that hold with high probability.
\begin{enumerate}
    \item[\textbf{Case $\delta<\Delta$}:] If none or only one change exist both sets will stay empty. Otherwise $\hat{\mathfrak{A}}_N(k)$ will be empty while $\tilde{\mathfrak{A}}_N(k)$ will always contain at least one element. The test with $\hat{\mathfrak{A}}_N(k)$ will therefore reject less often (asymptotically neither of them will reject at all) as $\hat L_{n,1}(k)$ is 0 when using $\hat{\mathfrak{A}}_N(k)$ while it is positive when using $\tilde{\mathfrak{A}}_N(k)$. 
    \item[\textbf{Case: $\delta=\Delta$}] For all $k$ before the first change with $|\psi^{(1)}-\psi^{(i+1)}|=\Delta$ occurs we observe the same behaviour as in the previous case. If no further changes occur we observe the same behaviour as when $\delta<\Delta$. Otherwise we will have $\hat{\mathfrak{A}}_N(k)=\tilde{\mathfrak{A}}_N(k)$ after the next change and both tests will show approximately the same behaviour.
    \item[\textbf{Case: $\delta>\Delta$}] Here a few slightly different things can happen and we look at the case where the first change with $|\psi^{(1)}-\psi^{(i+1)}|\geq \Delta$ in fact satisfies $|\psi^{(1)}-\psi^{(i+1)}|>\Delta$ as an illustrative example. Immediately after this change happens we still have  $\hat{\mathfrak{A}}_N(k)=\emptyset$ while $\tilde{\mathfrak{A}}_N(k)$ is non-empty so that the bootstrap quantile associated to $\tilde{\mathfrak{A}}_N(k)$ is larger (by the same reason as in the case $\delta<\Delta$), leading to a longer delay time before the  test rejects.
\end{enumerate}
To summarize: $\tilde{\mathfrak{A}}_N(k)$ does not immediately discard all changes smaller than $\Delta$, opting to use a running maximum instead. This more conservative choice ensures that the resulting method does not rely on knowledge of $\Delta$. This comes the cost of being more conservative as we may not deem past changes as uninformative anymore based on whether or not they exceed $\Delta$.
\putbib
\end{bibunit}


\begin{thebibliography}{}

\bibitem[Aue et~al., 2006]{aue:horvath:huskova:kokoszka:2006}
Aue, A., Horv{\'a}th, L., Hu{\v{s}}kov{\'{a}}, M., and Kokoszka, P. (2006).
\newblock Change--point monitoring in linear models with conditionally heteroskedastic errors.
\newblock {\em Econometrics Journal}, 9:373--403.

\bibitem[Aue and Horváth, 2013]{aue:horvath:2012}
Aue, A. and Horváth, L. (2013).
\newblock Structural breaks in time series.
\newblock {\em Journal of Time Series Analysis}, 34(1):1--16.

\bibitem[Aue et~al., 2009]{aue:horvath:reimherr:2009}
Aue, A., Horváth, L., and Reimherr, M.~L. (2009).
\newblock Delay times of sequential procedures for multiple time series regression models.
\newblock {\em Journal of Econometrics}, 149(2):174--190.

\bibitem[Aue and Kirch, 2024]{aue:kirch:2024}
Aue, A. and Kirch, C. (2024).
\newblock The state of cumulative sum sequential change point testing seventy years after {P}age.
\newblock {\em Biometrika}, Volume 111(2):367–391.

\bibitem[Bastian et~al., 2024]{bastian:basu:dette:2024}
Bastian, P., Basu, R., and Dette, H. (2024).
\newblock {Multiple change point detection in functional data with applications to biomechanical fatigue data}.
\newblock {\em The Annals of Applied Statistics}, 18(4):3109 -- 3129.

\bibitem[Bastian and Dette, 2025]{bastian:dette:2025}
Bastian, P. and Dette, H. (2025).
\newblock Multiscale detection of practically significant changes in a gradually varying time series.
\newblock {\em arXiv preprint: https://arxiv.org/abs/2504.15872}.

\bibitem[Berkes et~al., 2004]{berkes:gombay:horvath:kokoszka:2004}
Berkes, I., Gombay, E., Horv{\'a}th, L., and Kokoszka, P. (2004).
\newblock Sequential change-point detection in {GARCH}($p,q$) models.
\newblock {\em Econometric Theory}, 20:1140--1167.

\bibitem[Berkes et~al., 2011]{berkes:hoermann:schauer:2011}
Berkes, I., H\"ormann, S., and Schauer, J. (2011).
\newblock Split invariance principles for stationary processes.
\newblock {\em The Annals of Probability}, 39(6):2441--2473.

\bibitem[Cho and Kirch, 2024]{cho:kirch:2024}
Cho, H. and Kirch, C. (2024).
\newblock Data segmentation algorithms: Univariate mean change and beyond.
\newblock {\em Econometrics and Statistics}, 30:76--95.

\bibitem[Chu et~al., 1996]{chu:stinchcombe:white:1996}
Chu, C.-S.~J., Stinchcombe, M., and White, H. (1996).
\newblock Monitoring structural change.
\newblock {\em Econometrica}, 64:1045--1065.

\bibitem[Cryer et~al., 2009]{Cryer2009}
Cryer, P.~E., Axelrod, L., Grossman, A.~B., Heller, S.~R., Montori, V.~M., Seaquist, E.~R., and Service, F.~J. (2009).
\newblock Evaluation and management of adult hypoglycemic disorders: An endocrine society clinical practice guideline.
\newblock {\em The Journal of Clinical Endocrinology \& Metabolism}, 94(3):709--728.

\bibitem[Dehling, 1983]{dehling:1983}
Dehling, H. (1983).
\newblock Limit theorems for sums of weakly dependent {B}anach space valued random variables.
\newblock {\em Zeitschrift für Wahrscheinlichkeitstheorie und verwandte Gebiete}, 63:393--432.

\bibitem[Dette et~al., 2020a]{dette:eckle:vetter:2020}
Dette, H., Eckle, T., and Vetter, M. (2020a).
\newblock Multiscale change point detection for dependent data.
\newblock {\em Scandinavian Journal of Statistics}, 47(4):1098--1125.

\bibitem[Dette and Gösmann, 2020]{dette:gossmann:2019}
Dette, H. and Gösmann, J. (2020).
\newblock A likelihood ratio approach to sequential change point detection for a general class of parameters.
\newblock {\em Journal of the American Statistical Association}, 115(531):1361--1377.

\bibitem[Dette et~al., 2020b]{dette:kokot:volgushev:2020}
Dette, H., Kokot, K., and Volgushev, S. (2020b).
\newblock Testing relevant hypotheses in functional time series via self-normalization.
\newblock {\em Journal of the Royal Statistical Society Series B: Statistical Methodology}, 82(3):629--660.

\bibitem[Eichinger and Kirch, 2018]{eichinger:kirch:2018}
Eichinger, B. and Kirch, C. (2018).
\newblock A mosum procedure for the estimation of multiple random change points.
\newblock {\em Bernoulli}, 24:526--564.

\bibitem[Fremdt, 2015]{fremdt:2015}
Fremdt, S. (2015).
\newblock Page’s sequential procedure for change-point detection in time series regression.
\newblock {\em Statistics}, 49(1):128--155.

\bibitem[Frick et~al., 2014]{frick:munk:sieling:2014}
Frick, K., Munk, A., and Sieling, H. (2014).
\newblock Multiscale change point inference.
\newblock {\em Journal of the Royal Statistical Society, Series B}, 76:495--580.

\bibitem[Fryzlewicz, 2014]{fryzlewicz:2014}
Fryzlewicz, P. (2014).
\newblock Wild binary segmentation for multiple change-point detection.
\newblock {\em The Annals of Statistics}, 42(6):2243--2281.

\bibitem[Fryzlewicz, 2024]{fryzlewicz:2024}
Fryzlewicz, P. (2024).
\newblock Narrowest significance pursuit: Inference for multiple change-points in linear models.
\newblock {\em Journal of the American Statistical Association}, 119(546):1633--1646.

\bibitem[G{\"o}smann et~al., 2022]{gosmann:stoehr:heiny:dette:2022}
G{\"o}smann, J., Stoehr, C., Heiny, J., and Dette, H. (2022).
\newblock {Sequential change point detection in high dimensional time series}.
\newblock {\em Electronic Journal of Statistics}, 16:3608--3671.

\bibitem[Horv{\'a}th et~al., 2003]{horvath:kokoszka:huskova:steinebach:2003}
Horv{\'a}th, L., Hu{\v{s}}kov{\'{a}}, M., Kokoszka, P., and Steinebach, J. (2003).
\newblock Monitoring changes in linear models.
\newblock {\em Journal of Statistical Planning and Inference}, 126:225--251.

\bibitem[Hu, 2025]{Hu2025}
Hu, Q. (2025).
\newblock Testing relevant hypotheses in functional variance function via self-normalization.
\newblock {\em Scandinavian Journal of Statistics}, 52(3):1301--1329.

\bibitem[Jandhyala et~al., 2013]{jandhyala:2013}
Jandhyala, V., Fotopoulos, S., MacNeill, I., and Liu, P. (2013).
\newblock Inference for single and multiple change-points in time series.
\newblock {\em Journal of Time Series Analysis}, 34(4):423--446.

\bibitem[Kirch and Stoehr, 2022a]{kirch:stoehr:2022}
Kirch, C. and Stoehr, C. (2022a).
\newblock Asymptotic delay times of sequential tests based on u-statistics for early and late change points.
\newblock {\em Journal of Statistical Planning and Inference}, 221:114--135.

\bibitem[Kirch and Stoehr, 2022b]{kirch:stoehr:2022b}
Kirch, C. and Stoehr, C. (2022b).
\newblock Sequential change point tests based on u-statistics.
\newblock {\em Scandinavian Journal of Statistics}, 49(3):1184--1214.

\bibitem[Kitabchi et~al., 2009]{Abbas2009}
Kitabchi, A.~E., Umpierrez, G.~E., Miles, J.~M., and Fisher, J.~N. (2009).
\newblock Hyperglycemic crises in adult patients with diabetes.
\newblock {\em Diabetes Care}, 32(7):1335--1343.

\bibitem[Kutta and Dörnemann, 2025]{kutta:doernemann:2025}
Kutta, T. and Dörnemann, N. (2025).
\newblock Monitoring time series with short detection delay.
\newblock {\em Electronic Journal of Statistics}, 19(1):2239--2275.

\bibitem[Kutta et~al., 2024]{kutta:jach:kokoszka:2024}
Kutta, T., Jach, A., and Kokoszka, P. (2024).
\newblock Monitoring panels of sparse functional data.
\newblock {\em J. Time Series Anal.}

\bibitem[Lai, 1995]{lai:1995}
Lai, T.~L. (1995).
\newblock Sequential changepoint detection in quality control and dynamical systems.
\newblock {\em Journal of the Royal Statistical Society. Series B (Methodological)}, 57(4):613--658.

\bibitem[Lai, 2001]{lai:2001}
Lai, T.~L. (2001).
\newblock Sequential analysis: Some classical problems and new challenges.
\newblock {\em Statistica Sinica}, 11(2):303--408.

\bibitem[Mathew et~al., 2023]{mathew2020blood}
Mathew, T.~K., Zubair, M., and Tadi, P. (2023).
\newblock Blood glucose monitoring.
\newblock StatPearls Publishing.

\bibitem[Messer et~al., 2018]{messer:albert:schneider:2018}
Messer, M., Albert, S., and Schneider, G. (2018).
\newblock The multiple filter test for change point detection in time series.
\newblock {\em Metrika}, 81:589--607.

\bibitem[Niu et~al., 2016]{Niuetal2016}
Niu, Y.~S., Hao, N., and Zhang, H. (2016).
\newblock Multiple change-point detection: A selective overview.
\newblock {\em Statistical Science}, 31(4):611--623.

\bibitem[Pouwer and Hermanns, 2009]{pouwer2009insulin}
Pouwer, F. and Hermanns, N. (2009).
\newblock Insulin therapy and quality of life. a review.
\newblock {\em Diabetes/metabolism research and reviews}, 25(S1):S4--S10.

\bibitem[Siegmund, 1985]{siegmund:1985}
Siegmund, D. (1985).
\newblock {\em Sequential {A}nalysis: {T}ests and {C}onfidence {I}ntervals}.
\newblock Springer.

\bibitem[Truong et~al., 2020]{truongetal2020}
Truong, C., Oudre, L., and Vayatis, N. (2020).
\newblock Selective review of offline change point detection methods.
\newblock {\em Signal Processing}, 167:107299.

\bibitem[Venkatraman, 1992]{venkatraman:1992}
Venkatraman, E.~S. (1992).
\newblock {\em Consistency Results in Multiple Change-point Problems}.
\newblock PhD thesis, Stanford University.

\bibitem[Verzelen et~al., 2023]{verzelen:fromont:lerasle:reynaudbouret:2023}
Verzelen, N., Fromont, M., Lerasle, M., and Reynaud-Bouret, P. (2023).
\newblock Optimal change-point detection and localization.
\newblock {\em Annals of Statistics}, 51(4):1586--1610.

\bibitem[Vostrikova, 1981]{vostrikova:1981}
Vostrikova, L.~J. (1981).
\newblock Detecting ‘disorder’ in multidimensional random processes.
\newblock {\em Soviet Doklady Mathematics}, 24:55--59.

\bibitem[Wang et~al., 2020]{wang:yu:rinaldo:2020}
Wang, D., Yu, Y., and Rinaldo, A. (2020).
\newblock Univariate mean change point detection: Penalization, cusum and optimality.
\newblock {\em Electronic Journal of Statistics}, 14:1917--1961.

\bibitem[Woodall and Montgomery, 1999]{Woodall01101999}
Woodall, W.~H. and Montgomery, D.~C. (1999).
\newblock Research issues and ideas in statistical process control.
\newblock {\em Journal of Quality Technology}, 31(4):376--386.

\bibitem[Wu and Zhao, 2007]{Wu2007}
Wu, W.~B. and Zhao, Z. (2007).
\newblock Inference of trends in time series.
\newblock {\em Journal of the Royal Statistical Society: Series B (Statistical Methodology)}, 69(3):391--410.

\bibitem[Yoshihara, 1978]{yoshihara:1978}
Yoshihara, K. (1978).
\newblock Moment inequalities for mixing sequences.
\newblock {\em Kodai Math. J.}, 1:316--328.

\bibitem[Yu et~al., 2023]{Yu02102023}
Yu, Y., Padilla, O. H.~M., Wang, D., and Rinaldo, A. (2023).
\newblock A note on online change point detection.
\newblock {\em Sequential Analysis}, 42(4):438--471.

\end{thebibliography}


\begin{thebibliography}{}

\bibitem[Chevallier, 2023]{chevallier2023}
Chevallier, J. (2023).
\newblock Uniform in time modulus of continuity of brownian motion.
\newblock {\em ArXiv preprint: https://arxiv.org/abs/2312.15931}.

\bibitem[Kutta and Dörnemann, 2025]{kutta:doernemann:2025}
Kutta, T. and Dörnemann, N. (2025).
\newblock Monitoring time series with short detection delay.
\newblock {\em Electronic Journal of Statistics}, 19(1):2239--2275.

\end{thebibliography}
\end{document}